\documentclass{aa} 
\usepackage{graphicx}
\usepackage{txfonts}
\usepackage{lipsum}
\usepackage{subcaption} 
\usepackage{lscape}  
\usepackage{placeins}
\usepackage{natbib}  
\usepackage{epstopdf}
\usepackage{xcolor} 
\usepackage{multirow}
\usepackage[utf8]{inputenc}
\usepackage{array}
\usepackage{mathtools, nccmath, amsmath}
\usepackage{lscape}
\usepackage[version=4]{mhchem} 
\usepackage{soul}
\usepackage{siunitx}
\usepackage{hyperref}

\newcommand{\kms}{km~s$^{-1}$}
\newcommand{\vlsr}{$v_{\rm LSR}$}

\begin{document}

\title{Identification of molecular line emission using Convolutional Neural Networks}

\author{N. Kessler\inst{1} \and T. Csengeri\inst{1} \and D. Cornu\inst{2} \and S. Bontemps\inst{1} \and L. Bouscasse\inst{3}}

\institute{Laboratoire d'Astrophysique de Bordeaux, Univ. Bordeaux, CNRS, UMR 5804, F-33615 Pessac, France \and LERMA, Observatoire de Paris, PSL Research University, CNRS, Sorbonne Univ., UMR 8112, F-75014 Paris, France \and IRAM, 300 Rue de la Piscine, F-38046 Saint Martin d’Hères, France}

\date{Received  / Accepted }

\abstract
{Complex organic molecules (COMs) are observed to be abundant in various astrophysical environments, in particular toward star forming regions they are observed both toward protostellar envelopes as well as shocked regions. Emission spectrum 
especially of heavier COMs may consists of up to hundreds of lines, where line blending hinders  the analysis. However, identifying the molecular composition of the gas leading to the observed millimeter spectra is the first step toward a quantitative analysis. 
} 
{We develop a new method based on supervised machine learning to recognize spectroscopic features of the rotational spectrum of molecules in the 3mm atmospheric transmission band for a list of species including COMs with the aim to obtain a detection probability.}
{We used local thermodynamic equilibrium (LTE) modeling to build a large set of synthetic spectra of 20 molecular species including COMs with a range of physical conditions typical for star forming regions. We successfully designed and trained a Convolutional Neural Network (CNN) that provides detection probabilities of individual species in the spectra.}
{We demonstrate that the produced CNN-model has a robust performance to detect spectroscopic signatures from these species in synthetic spectra. We evaluate its ability to detect molecules according to the noise level, frequency coverage, and line-richness, and also test its performance for incomplete frequency coverage with high detection probabilities for the tested parameter space, and no false predictions. Ultimately, we apply the CNN-model to obtain predictions on observational data from the literature toward line-rich hot-core like sources, where detection probabilities remain reasonable with no false detection.
}
{We prove the use of CNNs facilitating the analysis of complex millimeter spectra both on synthetic spectra as well as first tests on observational data. Further analysis on explainability, as well as calibration using a larger observational dataset will help us to improve the performance of our method.}


\maketitle

\nolinenumbers

\section{Introduction}

Through the interplay of physical and chemical evolution of the interstellar medium, a variety of chemical species, including complex organic molecules (COMs), such as sugars, alcohols, aldehydes emerge \citep[see][for a review]{McGuire2022}. Their rotational (and in certain conditions vibrational) transitions in the (sub)millimeter range give access to these molecules in the gas phase. Based on decades of extensive observational efforts, the presence of chemical complexity across a broad range of astrophysical environments is well established. 
Numerous species have been identified toward expanding shells of evolved stars \citep[e.g.,][]{Kaminski2017}, as well as nearby  galaxies, where emission of COMs has also been confirmed \citep[e.g.,][]{Sewilo2018, Martin2021, Bouvier2025}. Sensitive observations reveal COMs in various  star forming environments, such as Galactic dense cores (pre-stellar cores, hot cores and hot corinos), shocks and extragalactic hot cores (for reviews \citealp[see e.g.,][]{Caselli2012, Jorgensen2020, McGuire2022, Ceccarelli2022, Shimonishi2023, Jimenez-Serra2025}).

To identify emission from molecules other than the most abundant \textsl{simple} species, multiple transitions need to be detected. Accurately assigning molecular transitions to observed spectral lines may be challenging, especially in chemically rich environments. This is because heavier COMs with large partition functions exhibit a plethora of (rotational) transitions due to their molecular structure, implying a  significant degeneracy in the potentially assigned transitions. Identifying the emission from such molecular species requires therefore an iterative fitting process using models of radiative transfer calculations, typically assuming local thermodynamic equilibrium (LTE) conditions. 
Modeling spectral line transitions over a range of upper-level energies ($E_{\rm up}$) enables  estimates of the physical conditions, with a particular emphasis on precise column densities required to accurately infer molecular abundances  \citep[cf.][]{Jorgensen2016, Mininni2020, Mercimek2022, Bouscasse2024, Belloche2025} enabling discussion of the chemical composition of astrophysical sources. Spectral surveys are particularly important in this aspect \citep[e.g.,][]{Caux2011, Lefloch2018, Bouscasse2024, Belloche2025, Moller2025}, as making use of a broad frequency coverage they provide numerous transitions from the same species.

While LTE modeling tools are available (e.g.,\,WEEDs, \citealp{Maret2011}; CASSIS, \citealp{Vastel2015}; XCLASS, \citealp{Moller2017}; MADCUBA, \citealp{Martin2019}; PySpecKit, \citealp{Ginsburg2011}; molsim \citealp{McGuire2024}), to obtain a proper model of the spectra, molecules at the origin of the observed spectral lines need to be first identified. Subsequent iterative minimization methods in a step-by-step approach provide the best fit results \citep[e.g.,][]{Moller2013, Qiu2025}.

Systematic analysis of a large number of spectra is therefore hindered by several factors. For detailed examples we refer to \citet{Bouscasse2022, Belloche2025, Moller2025}, who also provide an in-depth discussion of spectral line analysis methods. In short, first, the line-fitting process is iterative and delicate if species need to be first fitted individually. Second, a combination of limitations in instrumental setup, such as spectral resolution, and the source physical properties, such as thermal and turbulent line-width, are determinant for resolving individual spectral lines. Third, identifying and fitting emission from each species needs to take into account that the rotational spectra of COMs have several transitions that overlap in frequency even if individual spectral lines are resolved. This leads to line blending or line contamination, the former corresponding to completely overlapping spectral lines that are not separable without proper modeling, while the latter allows to identify the dominant line. Spectral confusion limit also needs to be considered, that is reached when emission from individual spectral lines cannot be distinguished due to a combination of the large number of spectral lines, and line overlap due to their turbulent and thermal line width, combined with unresolved source structure. For example, spectral confusion was an issue for Sgr~B2(N) \citep{Belloche2013}, however, resolving the source structure with ALMA reduced the intrinsic line widths, thereby mitigating spectral confusion \citep[c.f.][]{Belloche2022}. The physical structure of the source may also lead to emission in multiple velocity components, while temperature gradients and optical depth effects may result in  non-Gaussian spectral line profiles. Other instrumental effects, such as spectral artifacts, contamination by strong emission lines from the other side-band, and discontinuous frequency coverage add further complexity to the analysis.
 
Despite these challenges, new methods emerge to facilitate the task of analyzing COM emission in the spectrum. Matched filtering and stacking \citep{Loomis2018} help to increase the signal-to-noise to detect species \citep{Loomis2021, McGuire2022}, while Principal Component Analysis-based (PCA) filtering techniques may help to leverage complexity due to velocity components and line profiles \citep{Yun2023}. Automated mixture analysis exploits the chemical relevance of a molecule to facilitate the identification of species in chemical mixtures \citep{Fried2024}. 

This analytical approach can be complemented by data-driven approaches, particularly because machine learning methods have now been widely used for different thematics in the field of astrochemistry, such as for chemical modeling, reaction pathways, and computing binding energies \citep{Villadsen2022, Heyl2023, Behrens2024, Wang2025}, supervised and unsupervised machine learning techniques using vector representation of molecules have been developed to identify species with chemical similarity to those already detected in the ISM \citep{Lee2021}. Improving on this approach, \citet{Fried2023} and \citet{Scolati2023} include isotopologs and by coupling to LTE modeling they predict molecular column densities. Successful applications have been demonstrated to infer the chemical inventories of IRAS16293 \citep{Fried2023}, Orion-KL \citep{Scolati2023} and TMC-1 \citep{Toru2025}. Using different machine learning methods on radiative transfer models, \citet{Mendoza2025} extract information based on the line profiles of HCN and HNC molecules, while coupling chemical reaction calculations with neural network. Another approach has been to use information field theory to infer which lines trace best specific conditions in the ISM helping observational campaigns \citep{Einig2024}, while \citet{Grassi2025} couple 1D collapse models to thermochemical modeling to connect the model properties to specific tracers.

In other domains of astronomical spectroscopy, artificial neural networks (ANNs) have been efficiently used for X-ray \citep{Gonzalez2014, Dupourque2024} and optical spectroscopic datasets \citep[e.g.,][]{Bailer-Jones1997, Guiglion2024}. Motivated by this, here we aim to develop a neural network based approach to facilitate the analysis of complex spectroscopic (sub)millimetric data by significantly speeding up the line identification process in the millimeter spectral range providing a quasi-immediate prediction on which COMs may be present in the spectrum. To our knowledge this approach that has not yet been addressed in the literature.

We aim to design a multi-label classification method to detect and identify the spectral signature of other simple molecules as well as COMs within millimeter spectra with a focus on Galactic star forming regions, specifically hot core and hot corino-like environments. Although our main goal is to tackle identification of abundant COMs, it is necessary to consider in our modeling other simple organic and inorganic species since their emission may blend with strong transitions of COMs, hindering their detection. Unlike  the approach of e.g., \citet{Lee2021, Toru2025}, our methodology does not rely on underlying chemical models and as such the choice of the investigated species is arbitrary. Our objective is to demonstrate that a neural network can be trained and used to effectively discern the molecular composition of millimeter spectroscopic data focusing on the most abundant COM species exhibiting numerous transitions in their rotational spectrum. 

The applicability of our method depends on the physical conditions and source types represented in the training set. We also include isotopologs and assume LTE conditions that is a commonly accepted analysis method for hot core-like sources (\citealp{Rolffs2011, Giannetti2025}, e.g., \citealp{Al-Edhari2017, Duronea2019} for HC$_3$N). While we demonstrate the applicability of our method using data from the IRAM 30m telescope, our approach is designed to be independent of the observing setup.

The paper is organized as follows. We describe the construction of the training set in Sect.\,\ref{section:training_set} and present the used convolutional neural network (CNN) architecture, the training and its validation in Sect.\,\ref{section:cnn}. We demonstrate the capabilities and discuss the performance of the CNN-model in Sect.\,\ref{section:results}. Application to observational data is presented in Sect.\,\ref{section:obs_application}.

\section{Construction of the training set}
\label{section:training_set}

Supervised machine learning methods rely on labeled training sets to learn the underlying data distribution. For millimeter spectra, however, we lack sufficiently large sets of labeled observational datasets that would be usable for training, hence we need to rely on synthesized data where the physical parameters are well established and their labeling is unambiguous. However, the challenge is that such synthetic spectra represent several biases, most importantly they are incomplete in terms of molecular richness, may not represent well the source structure, and lack potential observational artifacts.

We use LTE modeling to obtain synthetic spectra of a list of species we will be focusing on. This approach allows us to explore a wide-range of physical parameters and molecular compositions. The free parameters for our LTE models are the molecular column density ($N_{\rm{X}}$, where $X$ represents different species), excitation temperature ($T_{\rm{ex}}$), kinematics of the gas (the rest velocity, \vlsr, and line-width, $\Delta v$), and the size of the emission for each species of the medium. This leads to a system with a degree of freedom of $n\times5$ to model, where $n$ is the number of molecules to consider. To systematically explore such a parameter space over a large range of physical conditions is challenging. Hence, we limit our approach to a handful of molecules that are widely abundant and have numerous rotational transitions in the millimeter spectral range and define the physical parameter space to be representative of the conditions observed in star forming regions. 

We focus here on the rotational spectrum of molecules in a frequency range of 80--115~GHz that covers a substantial fraction of the 3~mm atmospheric transmission window. This frequency range is interesting since it has less severe line confusion compared to higher frequencies. We explicitly chose this frequency range to avoid the CO ($J=1-0$) line at 115.271\,GHz due to its ubiquitously complex line profile.

\subsection{Molecular composition}\label{sub:molecular_composition} 

Our aim here is to facilitate the identification of species where LTE modeling is necessary for their firm identification due to their large number of rotational transitions. Therefore, we focus on COMs that are among the most abundant species found toward star-forming regions, and COMs that exhibit more than 100 rotational transitions in the considered frequency range above an $A_{\rm ij}$ threshold and below an upper level energy of $E_{\rm up}/k<$500~K. These species are O-bearing COMs, such as CH$_3$CHO, CH$_3$COCH$_3$, CH$_3$OCH$_3$, CH$_3$OCHO, (CH$_2$OH)$_2$, C$_2$H$_5$OH, and N-bearing COMs, like C$_2$H$_3$CN, C$_2$H$_5$CN, C$_3$H$_7$CN, CH$_3$NH$_2$ and HC(O)NH$_2$. All of these species are frequently detected toward chemically rich regions, such as hot molecular cores, hot corinos and shocks \citep[cf.][]{Belloche2013, Jorgensen2020, Palau2017, Bouscasse2024, Vastel2024}. The selected list of species corresponds to the most abundant COMs in these regions, yet it remains an arbitrary choice. For simplicity, we do not include S-bearing COMs, whose abundances are typically one to two orders of magnitude lower than those of COMs with analogous chemical structures. \citep[e.g.,][]{Baek2022, Nazari2024}.

For the sake of similarity with observational spectra, we add to this list smaller COMs, such as CH$_3$OH and other abundant species, like CH$_3$CCH, CH$_3$CN,  HC$_3$N, H$_2$CS, t-HCOOH, CH$_2$NH, and NH$_2$CN. They have lower number of rotational transitions (between 4 and 71) within our investigated limits of $E_{\rm up}/k$ and $A_{\rm{ij}}$, and many of these species are also easily identified without LTE modeling. These molecules also provide a way to benchmark the capability of our method to identify emission from simpler species. We list the studied chemical species together with their number of rotational transitions in Table \ref{tab:database_aij}.

\begin{table*}[]
    \centering
    \caption{Summary of the studied molecules and reference for the used entry of spectroscopic databases.}
    \begin{tabular}{c|c|c|c|c|c||c|c}
    \hline
    \hline
      Molecule  & Formula & Database & Tag & Version & Update & log$_{10}(A_{ij}$) [s$^{-1}$] & Nb. transitions\\
      \hline
      Acetaldehyde & CH$_3$CHO & JPL & $44003$ & V3 & 2021/03 & $-6$ & 172 \\
      Acetone & CH$_3$COCH$_3$ & JPL & $58003$ & V1 & 2017/08 & $-5$ & 1018 \\
      Cyanamide & NH$_2$CN & JPL & $42003$ & V1 & 2021/03 & $-7$ & 39 \\
      Cyanoacetylene & HC$_3$N & CDMS & $51501$ & V1 & 2017/08 & $-5$ & 4 \\
      Dimethyl Ether & CH$_3$OCH$_3$ & CDMS & $46514$ & V2 & 2017/08 & $-7$ & 310  \\
      Ethanol & C$_2$H$_5$OH & CDMS & $46524$ & V1 & 2017/08 &  $-9$ & 616 \\
      Ethyl Cyanide & C$_2$H$_5$CN & CDMS & $55502$ & V2 & 2017/08 & $-8$ & 350\\
      $a$-Ethylene Glycol & $a$-(CH$_2$OH)$_2$ & CDMS & $62503$ & V1 & 2017/08 & $-7$ & 1115 \\
      $g$-Ethylene Glycol & $g$-(CH$_2$OH)$_2$ & CDMS & $62504$ & V1 & 2017/08 & $-7$ & 1474  \\
      Formamide & HC(O)NH$_2$& CDMS & $45512$ & V2 & 2017/08 & $-6$ & 119 \\
      Formic Acid & t-HCOOH & CDMS & $46506$ & V1 & 2019/06 & $-8$ & 37\\
      Methanol & CH$_3$OH & JPL & $32003$ & V3 & 2017/08 &  $-10$ & 71 \\
      Methyl Amine & CH$_3$NH$_2$ & JPL & $31008$ & V1 & 2017/08 & $-8$ & 125 \\
      Methyl Cyanide & CH$_3$CN & CDMS & $41505$ & V2 & 2019/06 & $-5$ & 11 \\
      Methylene Imine & CH$_2$NH & CDMS & $29003$ & V2 & 2019/12 & $-7$ & 32 \\
      Methyl Formate & CH$_3$OCHO & JPL & $60003$ & V1 & 2017/08 & $-6$ & 380 \\
      Propyl Cyanide & C$_3$H$_7$CN & CDMS & $69002$ & V1 & 2019/12 & $-7$ & 1534 \\
      Propyne & CH$_3$CCH & CDMS & $40502$ & V3 & 2017/08 & $-7$ & 11 \\
      Thioformaldehyde & H$_2$CS & CDMS & $46509$ & V3 & 2020/01 & $-7$ & 8 \\
      Vinyl Cyanide & C$_2$H$_3$CN & CDMS & $53515$ & V1 & 2014/09 & $-5$ & 429 \\
      \hline
    \end{tabular}
    \tablefoot{The number of transitions is given for the frequency range of 80-115 GHz, an upper level energy ($E_{\rm up}/k$) of $500$~K and threshold on the Einstein $A_{\rm{ij}}$ coefficients given in the corresponding column.}
    \label{tab:database_aij}
\end{table*}

\subsection{LTE models} \label{sub:models}
We computed LTE synthetic spectra for the $20$ molecules discussed in Sect.\,\ref{sub:molecular_composition} (and listed in Table \ref{tab:database_aij}) for the investigated frequency range using the XCLASS software \citep{Moller2017}. The spectroscopic information was taken from the Cologne Database for Molecular Spectroscopy (CDMS) \citep{Muller2005,Endres2016} and Jet Propulsion Laboratory (JPL) \citep{Pickett1998} line catalogs as listed in Table \ref{tab:database_aij}, for each species, respectively. Using LTE models is a commonly accepted approach to model emission from the majority of the here discussed COMs, firstly because toward the densest inner regions of hot cores and hot corinos LTE conditions are satisfied, but also because collisional rate coefficients are not systematically available for the heavier COMs in our sample\footnote{Collisional rate coefficients for COMs discussed here are available for \ce{CH3OH} and \ce{CH3OCHO} in the LAMDA database \citep{Schoier2005}.}. 

We fixed the source rest velocity (\vlsr) to zero and use a spectral resolution of $1$ MHz corresponding to $\sim3$\,\kms\, giving in total  35000 channels for the frequency range between 85 and 115~GHz. Since we work directly with the frequency information, our model is applicable to any spectra with the proper \vlsr\ correction applied (see App.\,\ref{appendix:application}). We explore a column density range ($N_{\rm X}$) up to seven orders of magnitude, typically between $10^{12}$ and $10^{19}$~{cm$^{-2}$},  corresponding to physical conditions commonly observed toward high-mass star-forming regions at scales up to a few thousands au \citep[for a review see][]{Jorgensen2020}. We sample the column density range ($N_{\rm X}$) on a logarithmic scale using 40 points, and as discussed in Sect.\,\ref{sub:composite_spectra}, for certain species we use a different parameter range. The column density for CH$_3$OH is set to $10^{14}$ -- $10^{20}$\,cm$^{-2}$ as this species is observed to be abundant in star-forming regions, while for complex cyanides we use a range of $10^{12}$ -- $10^{18}$\,cm$^{-2}$, and for rarer molecules (ethylene glycol, propyl cyanide) $10^{12}$ -- $10^{17}$\,cm$^{-2}$. 
We compute models using five values of excitation temperature, corresponding to 30, 50, 100, 150, and 300\,K, sampling both quiescent and heated protostellar environments. We sample the line widths as 1, 3, 5, 10, and 12 \kms. An important source of often poorly constrained parameter is the ratio between the source size and that of the telescope beam. This can only be constrained by mapping experiments that lead to spatially resolved measurements of the molecular emission. For distant ($>$1~kpc) high-mass star forming regions, interferometric observations are typically required to measure the emission size of COMs. Treating the ratio between the source size and that of the telescope beam as a free parameter ensures that the models represent a broad range of observing configurations corresponding to both single-dish and interferometric measurements. For this purpose we use a 3\arcsec\ beam size and adopt a source size of 0.15, 0.3, 1.5, 3 and 15\arcsec\ which corresponds to a range of source size over beam size between 0.05 and 5. This parameter range covers both interferometric observations with marginally or completely resolved source structures for both nearby or more distant regions (cf.\,\citealp{Belloche2020, Feng2015, Bonfand2017, Giese2024}, resp.), as well as unresolved emission typically corresponding to single-dish observations of more distant regions \citep[e.g.,][]{Widicus2012,WidicusWeaver2017, Bouscasse2024}.

For simplicity, we neglect continuum emission other than the cosmic background radiation. Neglecting the contribution from moderately strong continuum radiation is not expected to have an impact hindering the line identification. Overall, we compute $5000$ models per molecule. The LTE spectrum containing emission from multiple species is obtained assuming that the spectra are linearly additive that is a reasonable assumption for optically thin emission.

Transitions from molecular isotopologs may help the identification of their parent  species by providing additional spectral lines, that can be a significant support for observations with limited spectral bandwidth. Therefore, we also include in our LTE models the most abundant isotopologs of these species as listed in Table \ref{tab:complement_isotop_vibration}, which are modeled together with their main isotopologue implying that the physical parameters are the same. We take the standard local interstellar medium (ISM) values as fix ratios. Our results are robust against variations in isotopic fractionation, as verified \textit{a posteriori} under conditions typical of the Galactic Center  \citep[cf.][]{Humire2020}. We also take into account the lowest vibrationally excited states for \ce{CH3CN} as well as the ones of \ce{CH3OH} together with their isotopologs (cf. Table \ref{tab:complement_isotop_vibration}). The ratios of the vibrationally excited states to the ground ($\varv=0$) states are set to one.

\subsection{Composite spectra} \label{sub:composite_spectra} 
The training set is compiled from composite synthetic spectra that are generated by random linear combinations of LTE models from individual species. When adding the spectra of multiple species we introduce a jitter by randomly shifting the spectrum up to two channels for each species in both red-shifted and blue-shifted direction to take into account that emission from different species may not originate from the same gas. To ensure diversity in the training set, the physical parameters of the spectra (excitation temperature, line width, source size over beam size) are independently varied for all species. Isotopologues, modeled jointly with their main species and sharing the same parameters, always accompany the main species. We also vary the number of species and column densities used for the composite spectrum that is discussed in detail in Sect.\,\ref{sub:training_set}.

The inclusion of a thermal noise to the spectra is necessary to enable a meaningful application to observational data. Once the composite spectrum is created, we add a Gaussian noise where the standard deviation is sampled from a uniform distribution giving an overall noise level that varies between  0.2~mK and 280~mK. 

We also introduce additional features to the spectra with the aim to enhance the robustness of the ANN-model against unknown components and mimic potential observational effects. First, we add fake emission lines that are randomly distributed in frequency to mitigate the effect of emission from species not included in our model. These fake lines account for between  $5$\% and $10$\% of the total number of transitions within a given spectrum, and they have a brightness temperature randomly drawn between $10$~mK and $300$~K following a log-normal distribution. Second, we also add artificial negative Gaussian lines to the spectra corresponding to absorption features originating from either physical or instrumental effects. We randomly select the number of absorption components for 50\% of the training set, using between 1 and 10 Gaussian components with a mean line width of 2.75~MHz and amplitudes randomly sampled from a uniform distribution with values between 30\% and 50\% of the strongest line in the band.

We find that these adjustments to the training sample are crucial for our trained model to achieve good generalization. 
In particular, including a set of artificial lines in the spectra adds spectral noise (with positive and negative features) to the training set which allows the ANN to better learn signal from species to be identified. Injecting these different forms of noise unrelated to molecular emission helps the network to be more robust against unknown transitions and therefore to be more adapt to work on observational spectra (see Sect.\,\ref{section:obs_application}). 

A further step is to simplify the task of network training by masking a small frequency range around  transitions from the most abundant simple molecules listed in Table~\ref{tab:small_molecules}. This implies that we exclude these frequency ranges from the analysis. This is necessary, because emission from several of these simple molecules is quasi omni-present in observational spectra, where COMs can be searched for. However, many of these species are expected to have optically thick emission in their low-$J$ transitions and due to their high abundances they are also sensitive probes of the  gas kinematics leading to complex line profiles, where concrete examples are $^{13}$CO, HCN, HNC, CS, SO, and N$_2$H$^+$ lines. Consequently, to enable a meaningful comparison to observational spectra, we mask ten channels centered on the rest frequency of their transitions. The number of transitions of COMs falling within these masks is listed in Table \ref{tab:hidden_transitions_mask}, where we note that this eliminates typically $<$10\% of their transitions, except for CH$_3$CN, where a more significant blending is noticed due to blended transitions from the weakest, $^{15}$N isotopolog.

\begin{figure*}
    \centering
    \includegraphics[width=1\linewidth]{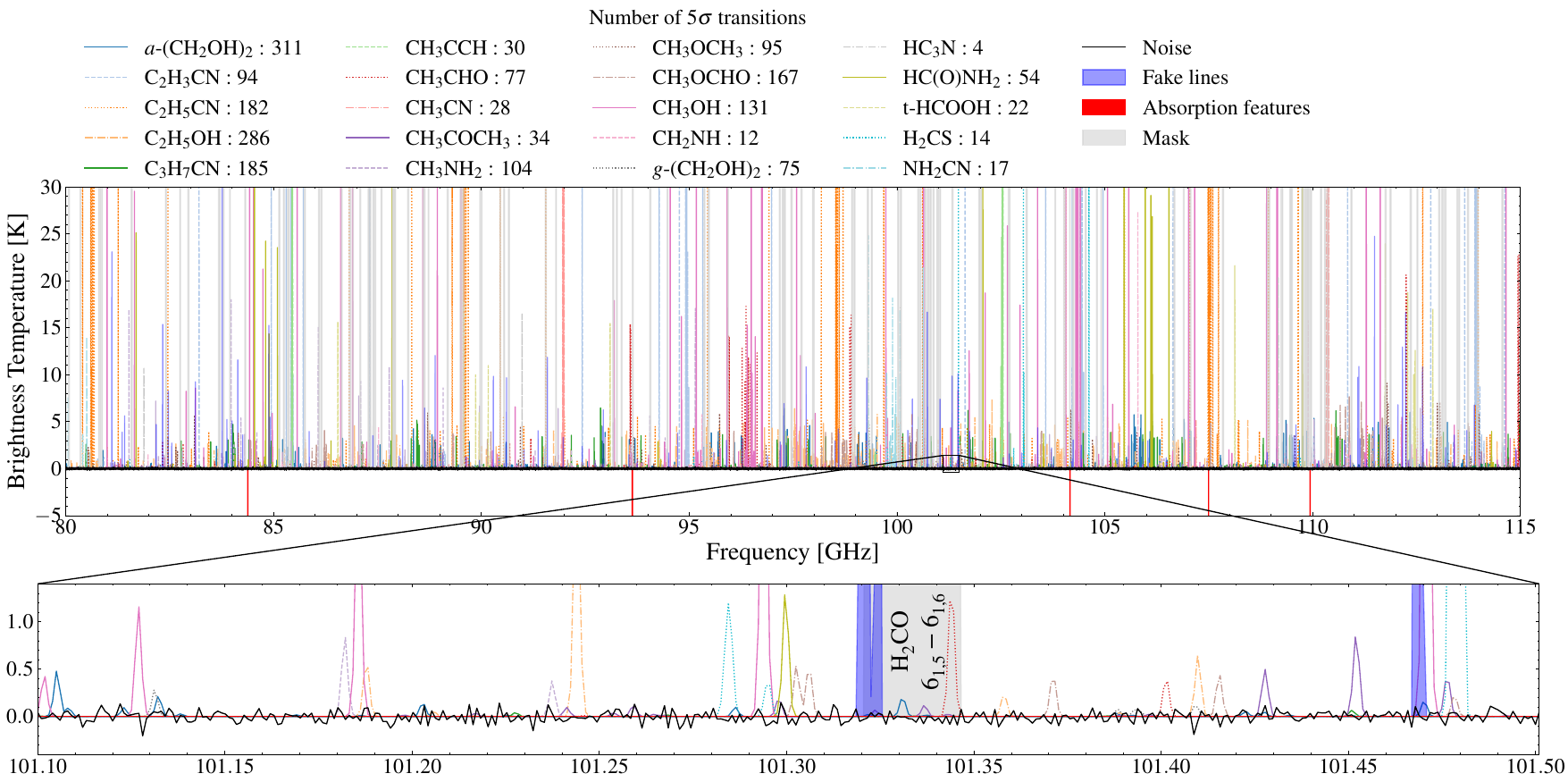}
    \caption{Synthetic spectrum of a classical hot core computed from Table \ref{tab:parameters_classical_hot_core} with a 5 \kms\ line width, a $50$\,mK Gaussian noise and a zoom on 400\,MHz. The LTE models are in colors. The mask computed with the molecules from Table \ref{tab:small_molecules} is in gray. The fake lines and absorption features are in blue and red respectively.}
    \label{fig:example_target}
\end{figure*}

Figure \ref{fig:example_target} shows an example resulting synthetic spectrum including emission from all investigated species with physical parameters corresponding to that of a typical hot core (see Table \ref{tab:parameters_classical_hot_core}) with lines having a width of 5 \kms, a noise level of 50\,mK assuming a source size that fills the 3\arcsec beam. As discussed above, our main objective is to develop a method applicable to chemically rich Galactic star forming regions, including hot cores, and show in Table \ref{tab:parameters_classical_hot_core} how representative our parameter range is for such sources.

\subsection{Constraining the training set} \label{sub:training_set} 

We create a training dataset of $4\times10^6$ composite synthetic spectra by randomly sampling the 5000 LTE model spectra for each of the 20 investigated species and all other parameters as described in Sect\,\ref{sub:models}. We define four sub-sets consisting of different molecular compositions, the first two follow a well-defined combination, the third is a random composition, and the fourth is a sub-set with just noise:

\begin{itemize}

\item The first sub-set of $10^6$ spectra aims to be representative of molecular environments with having at least 10 from the 20 randomly selected molecules in each spectrum with astrochemically reasonable conditions resembling that of hot corinos and hot cores. Certain molecules are expected to be chemically related, such as sharing common molecular precursors or having the same functional groups \citep{Garrod2006, Jorgensen2020}, and also observational results lend support to correlated abundance ratios among specific COMs \citep[e.g.,][]{Drozdovskaya2019, Coletta2020, Nazari2022, Bouscasse2024}. Here we take this into account by imposing  column density ratios separately for O- and N-bearing molecular families, as:
\begin{equation}
\frac{{N}_{\text{col}}(\text{O-bearing species})} {{N}_{\text{col}}(\text{CH}_3\text{OH})} \in [10^{-3},1].
\end{equation}
For N-bearing species that may become abundant in the hot gas phase, such as C$_2$H$_3$CN, C$_2$H$_5$CN, HC$_3$N and HC(O)NH$_2$:
\begin{equation}
\frac{{N}_{\text{col}}(\text{N-bearing species})}{{N}_{\text{col}}(\text{CH}_3\text{CN})} \in [10^{-3}, ~ 10^{2}].
\end{equation} 
For the rest of the species, such as NH$_2$CN, CH$_3$NH$_2$, CH$_2$NH, C$_3$H$_7$CN:
\begin{equation}
\frac{{N}_{\text{col}}(\text{N-bearing species})}{{N}_{\text{col}}(\text{CH}_3\text{CN})} \in [10^{-3}, ~ 10^{0}].
\end{equation} 
The column density for \ce{H2CS} and \ce{CH3CCH} remains randomly drawn from the global column density range.

We use here a broad range of abundance ratios that is globally consistent with observations \citep[cf.][]{Jorgensen2020, Belloche2025} and chemical model predictions \citep{Garrod2022}. We make the combination of the spectra in such a way to have species with weak lines ($a$- and $g$-\ce{(CH2OH)2}, \ce{C3H7CN}, \ce{CH3NH2}, and \ce{CH2NH}) equally well represented in the final sample. This dataset is called "recipe" for the rest of the paper. 

\item The second sub-set is also composed of $10^6$ spectra, it is the same as the previous one but with a random gap within each synthetic spectrum which has a width of $200$ to $800$ MHz. The detection status is modified if transitions in these gaps impact the detectability of each species. A risk during the training is that the network would only learn to recognize the few most obvious lines and neglect the others. Introducing these gaps forces the neural network to use less distinct attributes and to complete its information from other parts of the spectrum. Thus, this makes the model more efficient and robust to missing features. 

\item We create a sub-set of $4\times10^5$ spectra consisting of a random molecular composition and a noise distribution as previously described. The role of this dataset is to expand the diversity of the produced spectra. This dataset is referred to as "unconstrained" throughout the paper, as no constraints are applied to the molecular content or physical parameters, which can lead to chemically unrealistic combinations of spectra.

\item The last sub-set is composed of $1.6 \times 10^6$ spectra with only noise and artefacts so that the ANN can also learn the properties of these components. 

\end{itemize}

\subsection{Labeling}\label{sub:labeling}

The list of species used to obtain each combined synthetic spectrum provides a first initial labeling for the training set. However, thermal noise and the masking of transitions from simple, abundant species may reduce the detectability of the initially added molecules. Therefore, we revise the labeling to reflect the target values, which are determined by identifying the species that are detectable in each composite spectrum. We obtain the new labeling independently for each species by analyzing its individual spectrum prior to the linear combination, while still accounting for the thermal noise and  masking frequencies corresponding to transitions from abundant omitted species. For a species to be considered detectable we require that its spectrum contains at least two transitions with a peak intensity of $\geq5\sigma$, where $\sigma$ is the dispersion of the Gaussian noise distribution added to the spectra. If the spectrum of a molecule fulfills this detection criterion, it is flagged as detectable in the target vector of the corresponding composite spectrum.

We show the initial molecular composition of the "unconstrained" and "recipe" sub-sets of the synthetic spectra in Fig.\,\ref{fig:distrib_TS} as well as their revised labeling (constraining their target values). The obtained distribution for the molecular content on the whole training dataset is nearly uniform. The ANN sees between $10^6$ and $2\times10^6$ times each species during the training.

\subsection{Validation and test datasets}
We also create validation and test datasets each composed of $2\times10^4$ synthetic spectra independently drawn from the training set. The molecular composition of these spectra correspond to 50\% of our "recipe" and 50\% of our "unconstrained" approach. We use the test dataset in Sect. \ref{section:results} to evaluate the performance of the obtained ANN-model.

\section{Implementation of a CNN to learn spectral signatures} 
\label{section:cnn}

We aim to build a tool that extracts the molecular composition from a millimeter spectrum based on their rotational transitions. For this purpose we employ a classification model using a convolutional neural network architecture with supervised training \citep{LeCun2015}. In the following we discuss the implementation and training of this CNN.

\subsection{Input and normalization for the CNN}

\begin{figure*}
    \centering
    \includegraphics[width=1\linewidth]{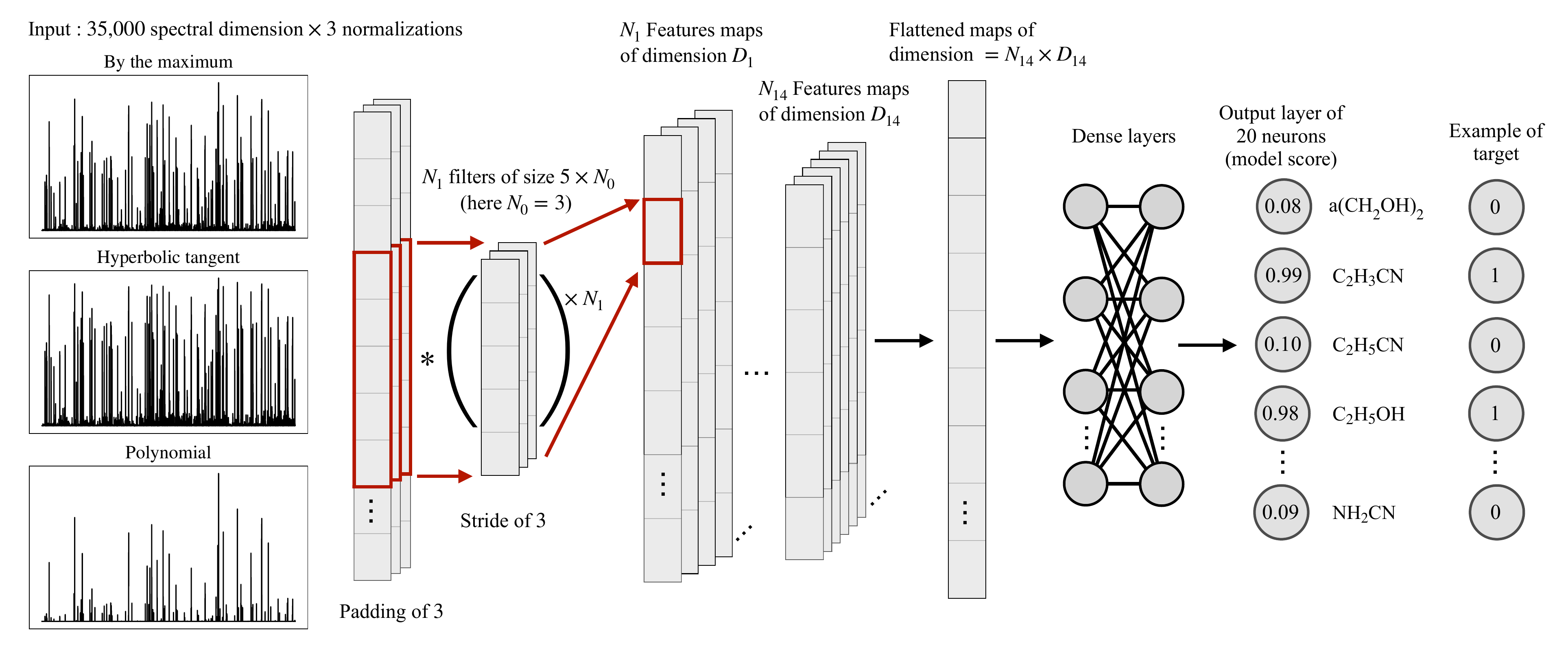}
    \caption{Scheme of the CNN architecture. The input data is an example of a composite spectrum according to the three normalizations, i.e., by the maximum (top), hyperbolic tangent (center), and polynomial (bottom). Filters are applied to the data to convolve the information and produce features maps. This operation is done for each of the convolutional layers. Dense layers then combine the extracted features and learn how to label the spectra depending on the provided target. The output layer is composed of one neuron per class giving a score between 0 and 1 independent between each other.}
    \label{fig:scheme_cnn}
\end{figure*}

To prepare the spectra for input into the neural network, we apply a normalization scheme widely used in machine learning applications, as it allows the CNN to focus on pattern recognition by getting rid of the data scale and thus facilitates training convergence. The broad dynamic range of line intensities in the spectrum makes normalization by its maximum value less effective for CNNs learning their signatures. Therefore, we provide three different normalizations of the training set to the CNN, including two that highlight respectively weak and strong transitions. As illustrated in Fig.\,\ref{fig:scheme_cnn}, the first transformation of the training set corresponds to a simple normalization, where each spectrum is normalized by its maximum value. The second transformation corresponds to a scaling where we compute the hyperbolic tangent of each spectrum multiplied by a factor $\alpha=1.5$, and then normalized by its maximum value, which increases the contrast of weak lines. For the third transformation we use a third degree polynomial of the normalized data to accentuate the features of strong lines. A single spectrum is therefore simultaneously injected to the network in these three forms of transformation.

Normalization is found to clearly influence the performance of  the CNN-model. The chosen methods were selected through testing to determine the most effective approach. In particular, we find that detection performances are systematically better when hyperbolic tangent normalization is combined with another one that preserves strong contrast in the maximum line intensity.

\subsection{Neural Network architecture}
For this study, we designed a custom network architecture, which allows for finer control over the reduction rate of the spectral dimension compared to using a classical backbone. The optimum  architecture was obtained empirically by starting with a very simple dense network and progressively increasing its depth and complexity. Systematic exploration of a gridded architectural space is very computationally expensive. Therefore, we used common substructures, setups, and possible adjunctions of layers as a baseline for our architecture and explored quite expansively the architectural space around that subspace. 

The architecture mostly follows the classical scheme of forward classification models with a convolution part to extract non-localized patterns, and then a dense part that recombines the obtained features to predict the output classes. Fig.\,\ref{fig:scheme_cnn} illustrates the structure of the 1D-CNN architecture. The convolutional layers were chosen to define the size of the receptive field at the end of the convolutional part to be of $8,786$ channels. It determines how the neural network filters and uses information from each single spectra \citep{Araujo2019}. Rotational transitions from asymmetric top molecules, such as most of the here investigated COMs, produce spectral features from the same molecule over the entire band, and hence we require the network to be able to process information from various parts of the spectra. We also include a few group normalization layers that improve the classification performance of our model \citep{Wu2018}. These normalizations help to correlate information from all scales in the convolution axis (here our 1D frequency axis) and also mitigate the risk of vanishing gradient issues. Like any layer normalization method, they also tend to speed up the training by reducing the number of iterations required to reach convergence.

Convolutional layers periodically apply a filter on their input according to a certain stride value. Thus, the network may give more importance to the signal coming from channels that fall on the central position of filters, and/or channels that are involved in the activation of several filters when there is superposition. This effect is reinforced for an application like ours where the position of the lines within the spectra is fixed, which can bias the network in its choice of important lines just by a geometric effect of the architecture. To avoid this, we add a jitter to the whole spectrum so that the peak of a line can be found on any element of the same filter. This amplitude is calculated as $- \left \lfloor \frac{k_{1}}{2}  \right \rfloor \leq x \leq  \left \lfloor \frac{k_{1}}{2}  \right \rfloor$  where $k_{1}$ is the size of the first filter, since it is the most sensitive to this effect. This also increments a translation invariance that could correspond to a potential error coming from the $v_{\rm{LSR}}$ estimation in observational data. We find that this addition tends to stabilize our results over slight architectural changes that was not supposed to affect much the network expressivity.

The convolutional part aims to extract coherent features in the spectra that are independent in their position in frequency space. These filters have a small number of parameters, but are applied to a large number of positions in the spectrum. Following a classical scheme, all the features identified by the convolutional part are then spatially recombined in the dense layers. As it is common, most of the model parameters are located in the dense layers due to the flattening of feature maps into a single 1D vector. The second dense layer has a $40$\% dropout rate which supports this aspect as more neurons need to be declared to retrieve the same expressiveness as without dropout. This reinforces the disparity in the number of parameters between the convolutional and the dense part. The dropout is constantly activated during the training as a mean of regularization to avoid overfitting and to build a more robust model capable of generalization \citep{Hinton2012}.

During inference, dropout is usually deactivated with the weights of the associated layers scaled down to compensate for the extra number of activated neurons. This approach averages all the possible models that random dropout selection could create, predicting a single averaged output vector in a single inference step. An alternative approach is to keep the dropout activated at inference time and perform multiple inferences for each input vector. This allows us to build a distribution of model scores (see Sect.\,\ref{sub:noise}) from which we can predict a mean or median value and a dispersion that reflects the uncertainty of the prediction. This second approach is often referred to as Monte-Carlo dropout \citep{Gal2015}. Unless specified we always use the prediction obtained through model averaging.

The output layer is a dense layer composed of $20$ neurons corresponding to the $20$ molecules to be detected. 
We use a logistic (sigmoid) activation to obtain an independent model score for each of the neurons from this output layer.
The stochasticity of the training will naturally push the network to predict a continuous value between 0 and 1 that will be proportional to its confidence in the presence of the molecule in a given input spectrum. Each class is represented by an output neuron independent from the activation of the other neurons of this final layer. 
In order to quantitatively interpret whether a given score value corresponds to a detection or non-detection of a molecule, we perform a "calibration" of the score values using the test dataset (see Sect.\,\ref{sub:calibration}).

Our final architecture is presented in Table \ref{tab:cnn} and has a total of $1,957,989$ parameters. It corresponds to the one presenting the best results while remaining computationally efficient. The network is implemented using the CIANNA (Convolutional Interactive Artificial Neural Networks by/for Astrophysicists) framework developed by \citet{Cornu2025}.

\begin{table*}[t]
    \caption{Detailed structure of the convolutional neural network for molecular detection as multi-label classification.}
    \centering
    \begin{tabular}{c|c|c|c|c|c|c|c}
    \hline \hline
     Id. & Type & Nb. filters & Filter size & Stride & Padding & Activation  & Spectral dimension \\
     \hline
    1 & Conv  & 16 & 5 & 3 & 3 &  RELU & 35,000 \\
    2 & Conv  & 16 & 3 & 3 & 1 &  RELU & 11,668\\
    3 & Conv  & 32 & 5 & 1 & 1 &  RELU & 3,890 \\
    4 & Conv  & 32 & 3 & 3 & 3 &  LIN. & 3,888 \\
    \hline 
    5 & Norm. & \multicolumn{4}{c|}{Group size : 1}  & RELU & 3,888\\
    \hline 
    6 & Conv  & 64 & 3 & 3 & 2 &  LIN. & 1,298 \\
    \hline 
    7 & Norm. & \multicolumn{4}{c|}{Group size : 2}  & RELU & 1,298\\
    \hline 
    8 & Conv  & 64 & 3 & 3 & 2 &  RELU & 434 \\
    9 & Conv & 64 & 3 & 3 & 2 &  LIN.  & 146 \\
    \hline 
    10 & Norm. & \multicolumn{4}{c|}{Group size : 2}  & RELU & 146\\
    \hline 
    11 & Conv  & 64 & 3 & 3 & 2 &  LIN. & 50 \\
    \hline 
    12 & Norm. & \multicolumn{4}{c|}{Group size : 2}  & RELU & 50\\
    \hline 
    13 & Conv  & 128 & 2 & 2 & 1 &  RELU & 18 \\
    14 & Conv  & 128 & 2 & 2 & 1 & RELU  & 10 \\
    \hline \hline
    Id. & Type  & Nb. neurons & \multicolumn{2}{c|}{Dropout} & \multicolumn{2}{c|}{Activation} & Input size \\
    \hline
    15 & Dense  & 1,024 & \multicolumn{2}{c|}{-} & \multicolumn{2}{c|}{RELU} & 769\\
    16 & Dense  & 1,024 & \multicolumn{2}{c|}{0.4} & \multicolumn{2}{c|}{RELU}  & 1,025\\
    17 & Dense & 20  & \multicolumn{2}{c|}{-} & \multicolumn{2}{c|}{LOGISTIC} & 1,025   \\
    \hline
    \end{tabular}
    \label{tab:cnn}
\end{table*}

\subsection{CNN training} 

The initialization of weights is Glorot normal \citep{Bengio2010}. CIANNA optimizer uses a mini-batch stochastic gradient descent with momentum. We set the batch size to 32, and the learning rate, $l_r$, follows an exponential decay to avoid oscillations around a minima or a too long convergence. It takes values from $l_{r,start} = 0.1$ to $l_{r,end} = 10^{-3}$ with a decay $d = 0.1$. We set the momentum to 0.6 which accelerates convergence and reduces the fluctuation of loss values during learning through gradient descent \citep{Qian1999}. We also add a small amount of weight decay $\lambda$ which is set at $5\times10^{-4}$. 

We use a loss function that is defined as the mean squared error (MSE) estimated on the validation dataset to update the weights of the network during the training. The under- and over-fitting is systematically checked during training with the help of the loss function computed 
over the validation dataset at each iteration (Fig.\,\ref{fig:training}). Once the training is done, the model having the lowest loss is taken for inference, which is also referred to as early stopping, to avoid using an overfitted model.

As discussed in Sect.\,\ref{section:training_set}, we generate sufficient simulated data (i.e., training set) to optimize its size for convergence, ensuring diversity so the network never sees the exact same spectrum twice. To handle this large sample size, we divide it into smaller portions, each corresponding to 1\% of the whole training dataset processed during one iteration. We then perform 100 iterations for the training. With $4\times10^6$ spectra in the training set and a computing precision of FP32C{\_}FP32A, the training takes 3.5~hours  on a server equipped with one NVIDIA A100 GPU. 

\section{Results and analysis} \label{section:results} 

\subsection{Metrics} \label{sub:metrics}

\begin{figure}[]
    \centering
    \includegraphics[width=1\linewidth]{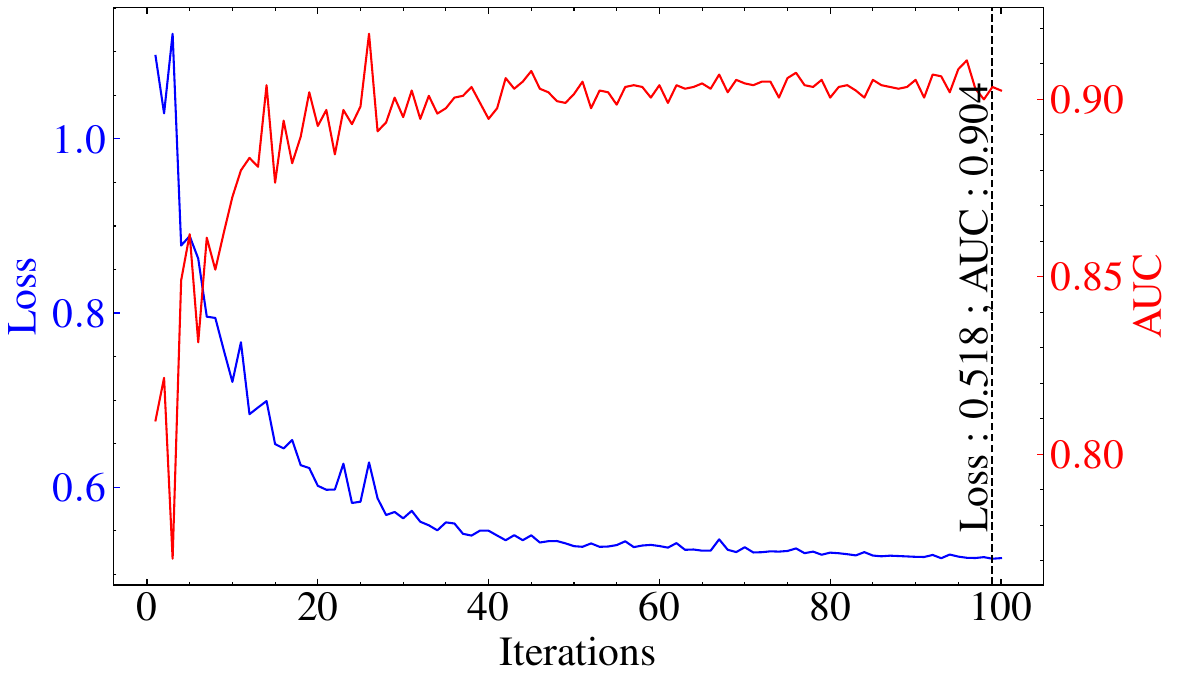}
    \caption{Loss function computed on the validation dataset and AUC values as a function of training iterations with specified minimum loss and the corresponding AUC.}
    \label{fig:training}
\end{figure}

We describe here the metrics to evaluate the classification performance of the CNN-model obtained during and after training. 
A first metric is the loss function that we use here on the test dataset. Figure \ref{fig:training} shows that the loss during the training decreases with the number of iterations. The MSE being an oversimplified metric, we also use other metrics such as the precision and recall that we calculate for each molecule as:
\begin{equation}
\text{Precision} =   \frac{\text{True detections}}{\text{True detections} + \text{False detections}}, 
\end{equation} and
\begin{equation} 
\text{Recall}  =    \frac{\text{True detections}}{\text{True detections} + \text{False non detections}}.
\end{equation} 

However, these metrics require the predictions of our model to be expressed as a per-class binary classification. This can be obtained by considering a species detected if its model score is above a certain threshold.

By sampling the per-species recall and precision for different model score thresholds over the whole test dataset we build a precision recall curve (or ROC curve) for each species (using a 0.01 threshold sampling step), which are represented in Fig.\,\ref{fig:roc_curves}. From this we also compute individual AUC (Area under the curve) values that are robust single value representations of the overall performances of the CNN-model on each species. To estimate the overall performance we also compute a mean AUC over all the species. The AUCs are computed during and after training to obtain a scalar value which allowed us to optimize the network architecture \citep{Bradley1997}. We represent the evolution of this mean AUC over training in Fig.\,\ref{fig:training} in comparison to the loss evolution demonstrating that the overall performance of the CNN-model improves over iterations.

The training state with the lower MSE is taken for model inference as AUC values fluctuate and may reach high values during the training. The here presented CNN-model is our best performing model with the minimum loss at iteration $99$ and a mean AUC over the molecules of $0.904$.

\subsection{Evaluation of the performance} \label{sub:performance}

\begin{figure}[]
    \centering
    \includegraphics[width=1\linewidth]{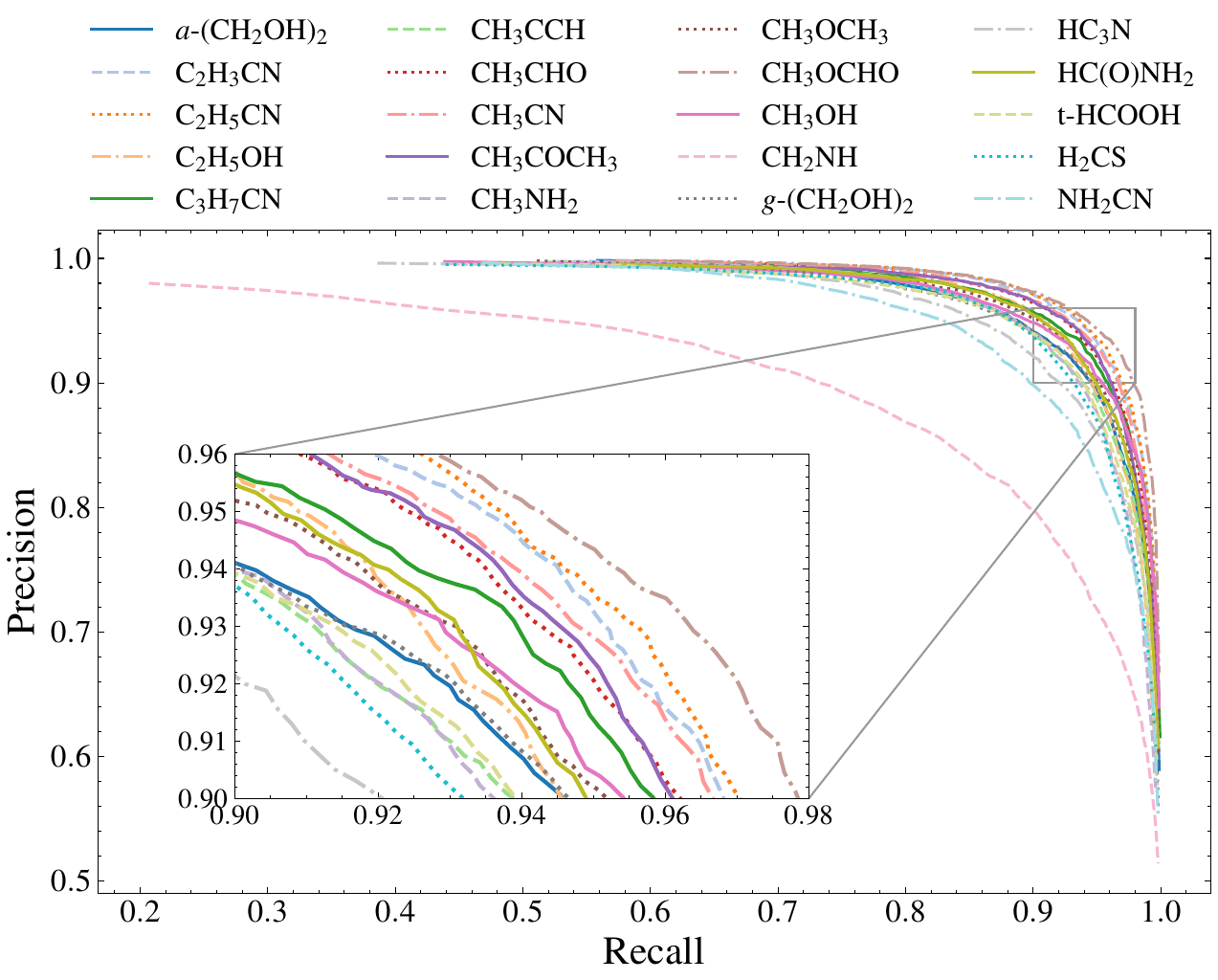}
    \caption{ROC curves of the molecules for which the CNN learned to detect their spectral signature. The values were computed on a [0, 1] range from the $x$ and $y$-axis.}
    \label{fig:roc_curves}
\end{figure}

The ROC curves show a simultaneously high recall and precision values for all species (Fig.\,\ref{fig:roc_curves}), with a somewhat lower values for CH$_2$NH. This allows us to conclude that the CNN-model has a generally high performance for the detection of each species with few false detections and missed detections.

\begin{figure}[]
    \centering
    \includegraphics[width=1\linewidth]{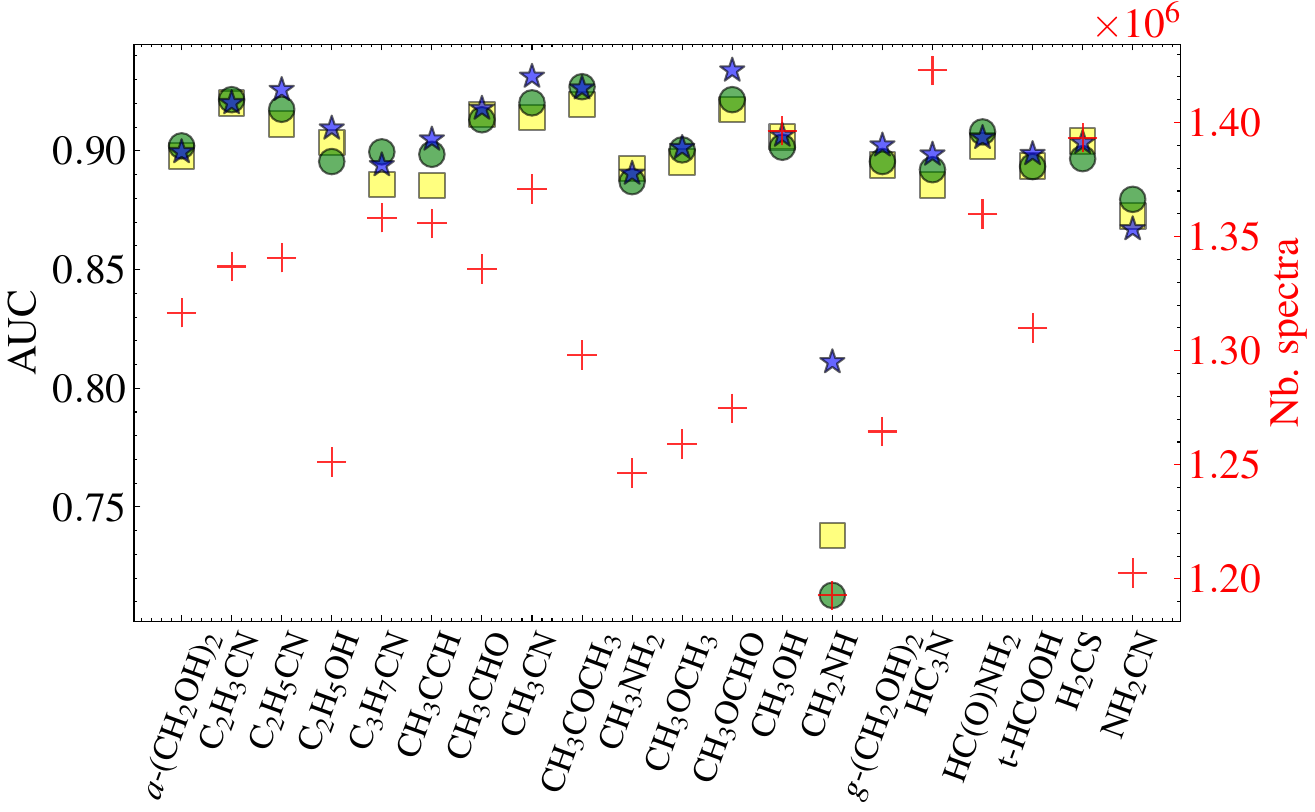}
    \caption{AUC as a function of molecules for three trainings of the multi-labeling CNN. The squares, triangles and stars are the AUC values. The red crosses correspond to the number of spectra where the molecules are detected.}
    \label{fig:reproductibility}
\end{figure}

We train our CNN-model three times with the exact same training setup but with different random starting weights in order to confirm the reproducibility of its results. In Fig.\,\ref{fig:reproductibility} we show the AUCs values for each species  for the three independent trainings and find negligible variations.
Their generally large ($>$0.89) values with a maximum dispersion of $0.04$ allow us to conclude that the CNN-model has a robustly high performance for the detection of each species. 
However, we find a lower mean AUC value (0.75) for \ce{CH2NH}, compared to the other species. We investigate whether this is related to the number of spectra containing the respective molecule in the training sample (red crosses in Fig.\,\ref{fig:reproductibility}), and find no clear correlation between the AUC values and the number of spectra with the given species.  
However, we find that one of the two bright transitions of \ce{CH2NH} is blended with the mask from Sect.\,\ref{sub:training_set}, meaning that this species is harder to detect unless the column density is high enough so the other transitions reach a sufficient signal-to-noise ratio (S/N).

\subsection{Impact of line density on the classifier performance} \label{sub:line_density}

We investigate the performance of the CNN-model as a function of line density using the spectra from the test dataset. For sources exhibiting a high molecular complexity, line density is a relatively straightforward metric to characterize spectra, and is applicable to both simulated and observational data. Here we measure the line density of the spectra using the \texttt{SciPy}\texttt{find\_peaks} function that uses local maxima above a fixed $5\sigma$ threshold to extract the number of detected transitions. The line density is obtained by dividing the number of lines  by our total frequency range of 35~GHz. 
 
\begin{figure*}[]
    \centering
    \includegraphics[width=1\linewidth]{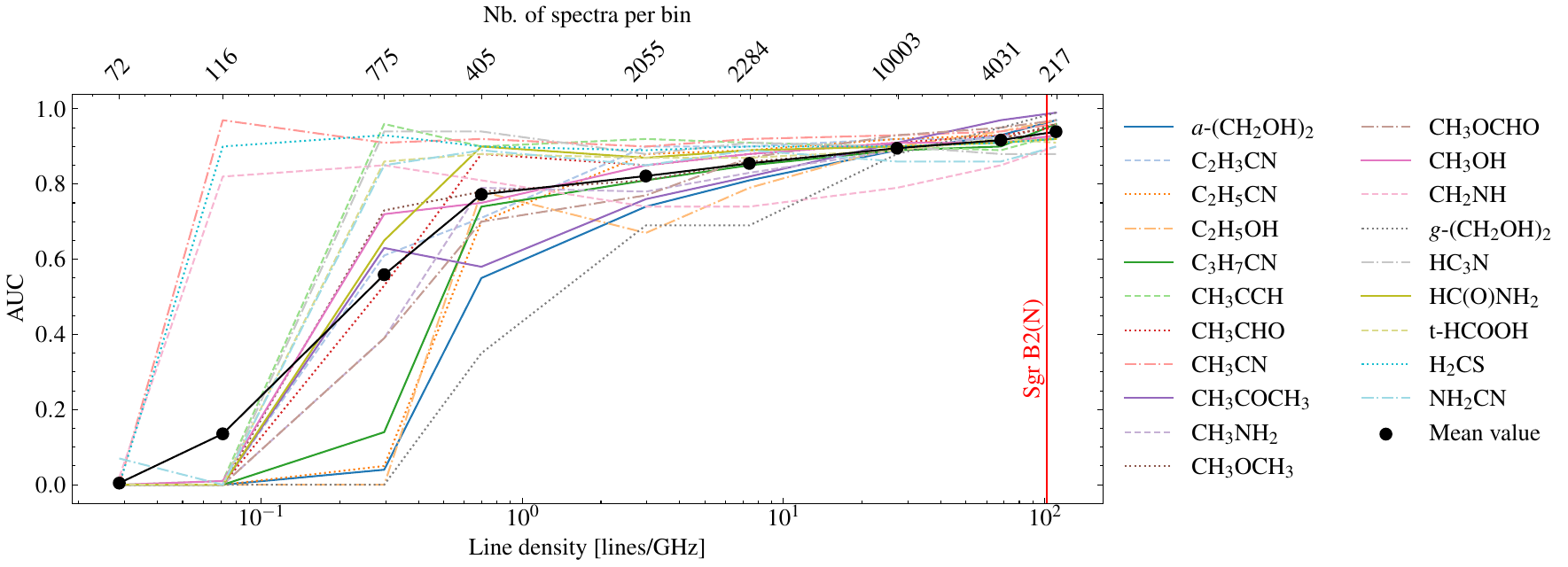}
    \caption{Performance of the model as the AUC in function of line density. In color: mean AUC values for individual molecules. Black : Mean global AUC value per bin. The number of spectra for AUC computation for each line density bin is specified. The line density of Sgr~B2(N) observed with the 30m telescope \citep{Belloche2013} is displayed as reference. }
    \label{fig:line_density_perf}
\end{figure*}

In Fig.\,\ref{fig:line_density_perf} we compute the AUCs as a function of line density for the spectra. The CNN-model shows a good performance for line-poor sources for molecules with only a few transitions, such as CH$_3$CN, H$_2$CS and CH$_2$NH with $11$, $8$ and $32$ transitions, respectively (cf. Table\,\ref{tab:database_aij}). We find that the mean AUC (of all species) increases with line density and from $2.97$ lines per GHz, it is systematically above a value of $0.8$ for 92.5\% of the spectra. We recall that the higher the AUC value, the better the performance. Our results, therefore, suggest that the CNN-model has a reliably high performance for a rather broad range of line densities. This behavior seems to indicate that the effect of line blending is marginal.

To put in context these line densities, we compare this range to that of the archetypal hot core, Sgr~B2(N). Observed with the 30m telescope, \citet{Belloche2013} finds a line density of $102$ lines per GHz for a frequency coverage of 79.990 -- 115.985\,GHz, while at higher angular resolution with ALMA \citet{Bonfand2017} find a line density of 438 to 460 for its most prominent hot cores over a similar frequency range of 84.1 - 114.4~GHz.  These frequency coverages are very similar to ours, however, their spectral resolution is higher by a factor of $\sim2-3$, hindering a direct comparison with these line density values. Resampling these spectra to a 1~MHz resolution would still preserve individual lines, however, line blending would likely decrease the number of lines assuming the same noise threshold. Furthermore, \citet{Bonfand2017} uses a different approach to estimate line density, their values are therefore only approximately comparable to the line densities we measure here. The largest value for the line density that we measure in our test set is $162$, while the mean and median line densities are $30$ and $23$, respectively. This suggests that the most extreme line rich sources are at the limit of our synthetic spectra, which is consistent with the fact that we consider here only the most abundant, and thus a limited number of molecules of only 20 species. Although such a stable and reliable performance for a broad range of line densities was one of our objectives, we caution that spectra with higher complexity (i.e., more lines) may not be well treated by our current CNN-model. A larger number of species must be considered in our model spectra in order to reach a line density similarly high to that of Sgr~B2(N).

\subsection{Calibration of the model score} \label{sub:calibration}

In the previous sections we evaluated the global statistical performance of the CNN-model based on the AUC which is an averaged measure of a varying threshold for the model score. However, the results of the CNN-model need to be interpreted for spectra where the target is not known. To do this, as described in Sect.\,\ref{sub:metrics}, the score given by the final output layer can be converted to either a logical output (detection versus non detection), or a detection probability. We find that the latter is more informative on the ability of the model to identify the spectral signatures of molecules, and therefore it is better suited for our applications. 

To obtain a detection probability, we need to calibrate the model score using the test dataset (see Sect.\,\ref{sub:training_set}).
We apply our CNN-model to the test dataset, that provides a list of detection scores between 0 and 1 for each spectrum, regardless of whether they contain the molecule or not. We then choose a score interval, for example $0.6 \pm 0.01$, for each species independently and select all the spectra with this model score. To convert this model score to a detection probability, we then evaluate the proportion of these selected spectra for which a detection of the species was expected, which can be expressed as:

\begin{equation} \label{eq:confidence}
    P_{\rm det} = \frac{\text{Nb. of true detections per bin}}{\text{Nb. of true det. + false det. per bin}}. 
\end{equation}

\begin{figure}
    \centering
    \includegraphics[width=1\linewidth]{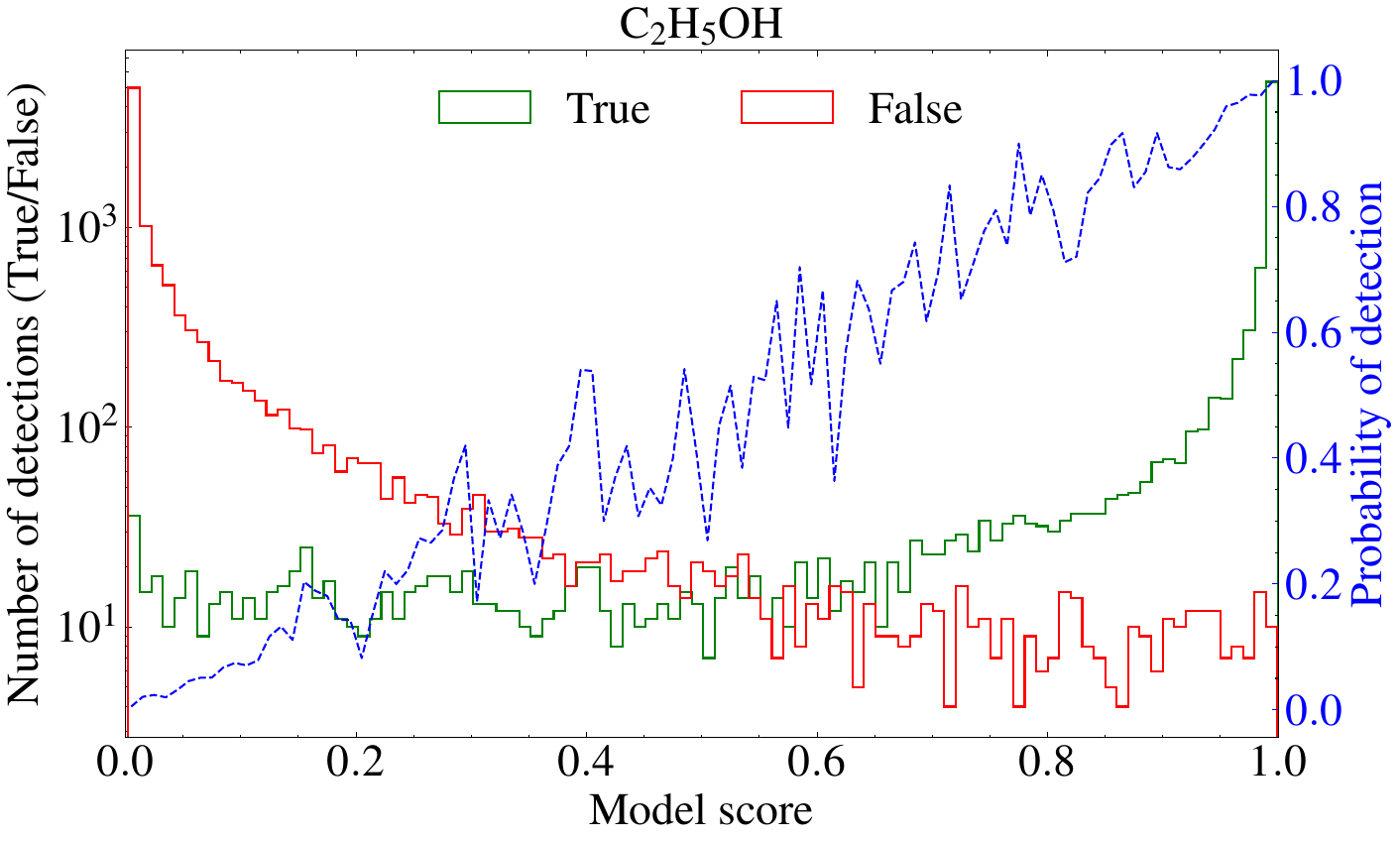}
    \caption{Calibration curve for C$_2$H$_5$OH based on the number of true and false detections over the test dataset.}
    \label{fig:cal_C2H5OH}
\end{figure}

By discretizing our [0,1] score interval we can compute a probability curve as a function of the predicted score for each species. We show an example obtained for the \ce{C2H5OH} molecule in Fig.\,\ref{fig:cal_C2H5OH}, where the detection probability scales roughly linearly with the model score. Based on our test dataset, we find a similar, roughly linear trend for all molecules. This linear scaling law is mostly the result of a relatively balanced training sample between the detectable and non-detectable cases. While this test confirms that we could use the detection score as a direct proxy for detection probability, this would not be the case for a different training set composition. In fact, the test set for calibrating the detection probability could be optimized if the range of physical conditions of the investigated source type is well defined. For the sake of consistency, we use the same calibration data set for all applications in this study. In the following, the value obtained by converting the model score through the calibration curve is called detection probability $P_{\rm det}$ and is expressed in percentage.

\subsection{Effect of the noise} \label{sub:noise}

\begin{figure}
    \centering
    \includegraphics[width=1\linewidth]{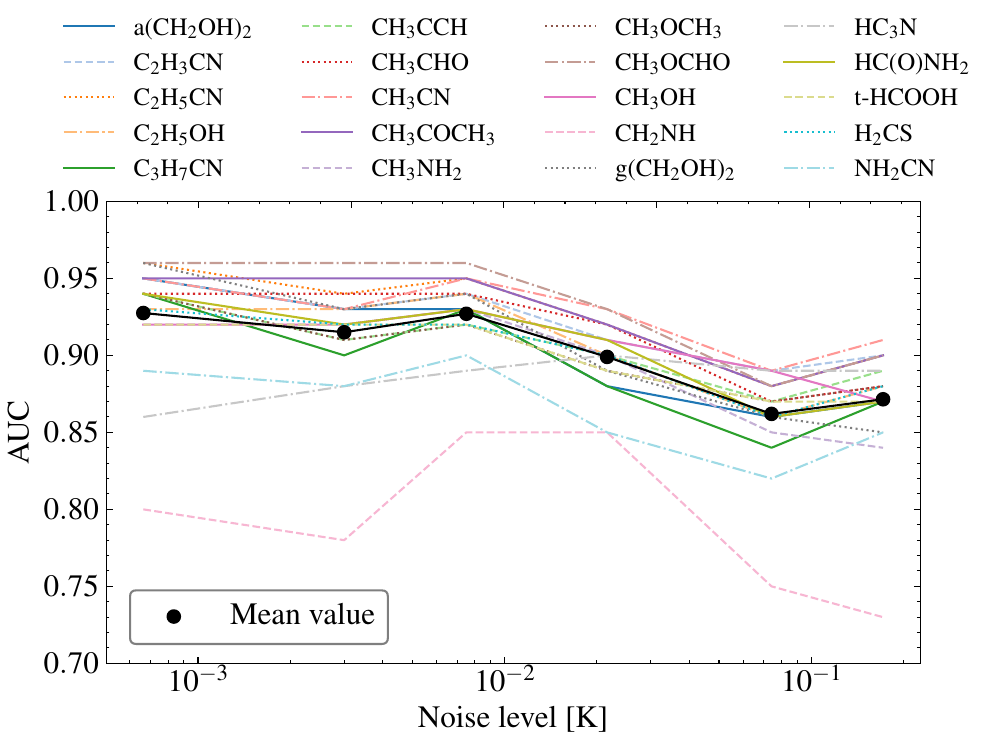}
    \caption{Performance of the model as a function of noise. The AUC for each molecule is in color, the averaged value over all species is in black.}
    \label{fig:AUC_noise}
\end{figure}

In our simplified synthetic spectra, we expect thermal noise to be the dominant factor affecting the detectability of species, although line blending may also further reduce detectability. However, since we explore a broad range of column densities and source size over beam sizes, the detectability of molecules is not expected to scale linearly with the noise. Here we explore the impact of noise globally on the results of the CNN-model by representing in Fig.\,\ref{fig:AUC_noise} the AUC for each molecule as a function of noise in the test dataset. We find that the noise starts to have a impact above 20\,mK resulting in a slight decrease in the AUCs, although the overall values remain still high above $0.85$. As discussed in Sect.\,\ref{sub:performance}, \ce{CH2NH} has systematically lower AUC values.

In order to investigate the impact of noise on detection probability, we use an example spectrum containing emission from all species (with the physical parameters listed in Table\,\ref{tab:parameters_classical_hot_core}, a 5 \kms\ line width and a source size equal to the beam size of 3\arcsec) without fake lines or absorption features, albeit with our initial mask from Sect.\,\ref{sub:composite_spectra}. We introduce three different noise levels of $100$\,mK, $500$\,mK, and $1$\,K, and extract the number of transitions with $\geq$ $5\sigma$,  as well as the signal-to-noise ratio of the weakest such line from the LTE spectrum of each individual molecule. This is an important metric to evaluate the results of the CNN-model, allowing us to compare our classical analysis approach with the CNN-model prediction that leverages the entire rotational spectrum. Although the two highest noise levels are outside the range used for the training dataset, since we use normalized spectra as input, we do not expect this to have an impact on the predictions.

The detection probability provided by the CNN-model is obtained through 100 realizations of MC-dropout on each of the three spectra. From these results, we take the median of the detection probabilities, and the median absolute deviation (MAD) as its uncertainty. In Table \ref{tab:SNR_study} we present the manually extracted number of transitions above 5$\sigma$, and the corresponding minimum signal-to-noise ratio of these lines, as well as the detection probability for each species provided by the CNN-model. This serves as a sanity check to compare the model performance with expectations from the manual analysis.

We have a very high median detection probability for the lowest, 100~mK noise level where most probabilities are above 99\%, and uncertainties $<$1\% (with a 5\% uncertainty for \ce{CH3COCH3}. The exception is $g$-\ce{(CH2OH)2}, where the detection probability is lower at $85$\,\% with uncertainties of 4-6\%. These results stress the excellent performance of the CNN-model on this highest single-to-noise, highly idealized spectrum. 
Increasing the noise by a factor of five, in this example spectrum we loose detectable transitions from \ce{CH3COCH3} and $g$-\ce{(CH2OH)2}. Although we would expect a detection probability of 0, the CNN-model gives a detection probability of $25.4$\,\% with a large dispersion of about 10\% for \ce{CH3COCH3}. For $g$-\ce{(CH2OH)2}, we obtain a low detection probability with a 2.9$^{+2.0}_{-1.4}$\,\% value. Thus, the results for $g$-\ce{(CH2OH)2} are consistent with our expectations, however for \ce{CH3COCH3} it is the significant median absolute deviation that suggests the non-detectability of the molecule. Considering the rest of the species, those impacted by the higher noise are $a$-\ce{(CH2OH)2}, \ce{CH3OCHO}, and \ce{CH2NH} for which the CNN-model gives a lower detection probability between 70 and 86 \%, with higher median absolute deviations reaching 4--12\%. Finally, with a $1$\,K noise, many COM lines get below the $5\sigma$ detection threshold. Interestingly, the detection probabilities do not strictly correlate with the number of detectable transitions. For example, while for CH$_2$NH, only $3$ detectable transitions remain, we obtain a $72.7$\,\% detection probability with an uncertainty of a few per cents. Similarly, \ce{C2H5OH} has only 11 detectable transitions but still presents a large probability with a median value of $81.6$\,\%. Whereas with $6$ detectable lines for $a$-(CH$_2$OH)$_2$, we obtain only $24.8$\,\% detection probability and a large uncertainty. Visual inspection of the spectrum suggests that only three of these transitions are not blended or masked, and the CNN-model uses two of them (see Sect.\,\ref{sub:features}) to identify this molecule. These two lines are, however, weak and close to the detection limit of 5$\sigma$. Despite seven detectable transitions, the detection probability for \ce{CH3OCHO} decreases to $38.5$\,\% with large errors. For the other species with a larger number of detectable transitions the detection probability still remains above $90$\,\% and with a negligible uncertainty.

Overall we find that the detection probability is impacted by increasing the noise. As expected, the CNN-model gives systematically the highest detection probabilities for species having $\geq2$ transitions at $\geq5\sigma$, that was imposed by our training set. We also find that the CNN-model performance clearly does not solely depend on the number of detectable transitions. Nevertheless, our CNN-model produces an excellent result for the example synthetic spectrum. In the following sections, we further evaluate the CNN-model's performance under various constraints on the input scenarios.

\begin{table*}[]
    \centering
    \caption{Monte Carlo dropout results for the impact of the noise level on the detection probability.}
    \begin{tabular}{c||c|c|c||c|c|c||c|c|c}
    \hline \hline
    \multirow{2}{*}{\textbf{Molecule}} & \multicolumn{3}{c||}{\textbf{Noise level : 100 mK}} & 
    \multicolumn{3}{c ||}{\textbf{Noise level : 500 mK}} & \multicolumn{3}{c }{\textbf{Noise level : 1 K}}\\
    \cline{2-10}
      & 5$\sigma$ lines & SNR$_{\text{min}}$ & $P_{\rm det}$[\%] & 5$\sigma$ lines & SNR$_{\text{min}}$ & $P_{\rm det}$[\%] & 5$\sigma$ lines & SNR$_{\text{min}}$ & $P_{\rm det}$[\%]\\
     \hline 

$a$-(CH$_2$OH)$_2$ & 158 & 5.03 & 99.9$^{+0.1}_{-0.3}$ & 61 & 5.08 & 70.5$^{+9.0}_{-12.0}$ & 6 & 5.12 & 24.8$^{+13.0}_{-8.0}$ \\
C$_2$H$_3$CN & 72 & 5.13 & 99.9$^{+0.1}_{-0.1}$ & 53 & 5.54 & 99.9$^{+0.0}_{-0.1}$ & 47 & 8.60 & 99.9$^{+0.0}_{-0.1}$ \\
C$_2$H$_5$CN & 150 & 5.01 & 99.8$^{+0.1}_{-0.1}$ & 92 & 5.56 & 99.8$^{+0.0}_{-0.1}$ & 52 & 5.06 & 99.8$^{+0.0}_{-0.1}$ \\
C$_2$H$_5$OH & 214 & 5.15 & 99.8$^{+0.1}_{-0.1}$ & 69 & 5.02 & 97.4$^{+1.0}_{-1.7}$ & 11 & 5.01 & 81.6$^{+5.0}_{-7.2}$ \\
C$_3$H$_7$CN & 92 & 5.12 & 99.8$^{+0.1}_{-0.1}$ & 68 & 5.18 & 99.4$^{+0.0}_{-0.7}$ & 10 & 5.03 & 93.2$^{+4.0}_{-2.1}$ \\
CH$_3$CCH & 27 & 6.77 & 99.7$^{+0.1}_{-0.1}$ & 10 & 21.74 & 99.6$^{+0.0}_{-0.3}$ & 10 & 10.84 & 99.6$^{+0.0}_{-0.3}$ \\
CH$_3$CHO & 45 & 5.06 & 99.8$^{+0.1}_{-0.1}$ & 28 & 5.33 & 99.2$^{+0.0}_{-1.5}$ & 18 & 8.41 & 96.5$^{+1.0}_{-2.0}$ \\
CH$_3$CN & 22 & 5.10 & 99.9$^{+0.1}_{-0.1}$ & 15 & 5.12 & 99.8$^{+0.0}_{-0.1}$ & 9 & 5.11 & 99.9$^{+0.0}_{-0.1}$ \\
CH$_3$COCH$_3$ & 9 & 5.57 & 97.5$^{+0.6}_{-5.0}$ & 0 & 0.00 & 25.4$^{+10.0}_{-8.7}$ & 0 & 0.00 & 17.6$^{+9.0}_{-4.7}$ \\
CH$_3$NH$_2$ & 87 & 5.04 & 99.7$^{+0.1}_{-0.1}$ & 29 & 5.05 & 99.7$^{+0.0}_{-0.1}$ & 16 & 5.22 & 99.7$^{+0.0}_{-0.1}$ \\
CH$_3$OCH$_3$ & 77 & 5.10 & 99.9$^{+0.1}_{-0.1}$ & 18 & 5.80 & 95.5$^{+1.0}_{-1.0}$ & 10 & 5.68 & 73.6$^{+7.0}_{-9.8}$ \\
CH$_3$OCHO & 109 & 5.03 & 99.8$^{+0.1}_{-0.1}$ & 51 & 5.12 & 82.4$^{+7.0}_{-12.2}$ & 7 & 5.03 & 38.5$^{+10.0}_{-11.3}$ \\
CH$_3$OH & 121 & 5.09 & 99.8$^{+0.1}_{-0.1}$ & 67 & 5.01 & 99.8$^{+0.0}_{-0.1}$ & 58 & 5.16 & 99.8$^{+0.0}_{-0.1}$ \\
CH$_2$NH & 10 & 9.73 & 97.9$^{+0.1}_{-0.1}$ & 5 & 5.38 & 86.1$^{+4.0}_{-4.0}$ & 3 & 5.72 & 72.7$^{+4.0}_{-5.9}$ \\
$g$-(CH$_2$OH)$_2$ & 5 & 5.13 & 86.4$^{+4.3}_{-6.5}$ & 0 & 0.00 & 2.9$^{+2.0}_{-1.4}$ & 0 & 0.00 & 1.0$^{+1.0}_{-0.3}$ \\
HC$_3$N & 4 & 106.39 & 99.4$^{+0.2}_{-0.3}$ & 4 & 21.33 & 97.0$^{+2.0}_{-1.2}$ & 4 & 10.64 & 91.8$^{+3.0}_{-5.1}$ \\
HC(O)NH$_2$ & 40 & 5.27 & 99.7$^{+0.1}_{-0.1}$ & 18 & 5.74 & 98.7$^{+1.0}_{-0.8}$ & 14 & 14.29 & 99.5$^{+0.0}_{-0.6}$ \\
t-HCOOH & 16 & 5.49 & 99.7$^{+0.1}_{-0.2}$ & 13 & 11.68 & 98.2$^{+1.0}_{-1.8}$ & 13 & 5.82 & 96.2$^{+1.0}_{-1.3}$ \\
H$_2$CS & 12 & 7.53 & 99.5$^{+0.1}_{-0.1}$ & 4 & 20.59 & 99.5$^{+0.0}_{-0.2}$ & 4 & 10.27 & 98.8$^{+0.0}_{-1.1}$ \\
NH$_2$CN & 17 & 10.91 & 99.7$^{+0.1}_{-0.1}$ & 14 & 5.78 & 99.2$^{+0.0}_{-0.4}$ & 11 & 5.43 & 97.9$^{+0.0}_{-0.7}$ \\

   \hline
   \end{tabular}
    \label{tab:SNR_study}
\end{table*}

\subsection{False positives} \label{sub:false_positives}

\begin{figure*}
    \centering
    \includegraphics[width=1\linewidth]{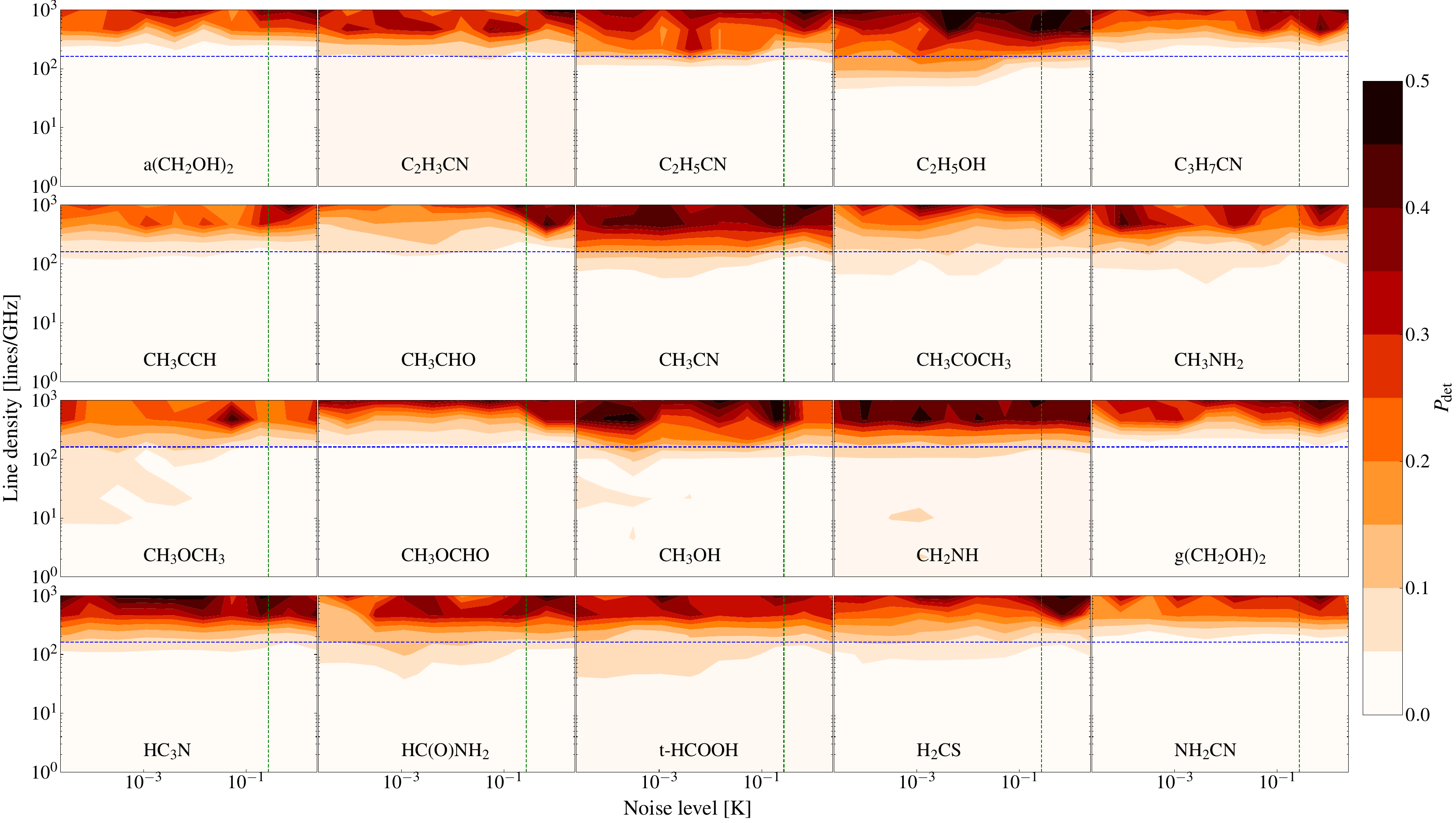}
    \caption{Detection probability (average on 1000 realizations) as a function of the line density and the noise level. The blue horizontal line and the green vertical line correspond to the line density and noise limits of our training dataset. }
    \label{fig:hallu}
\end{figure*}

To test to which extent our model is susceptible to hallucination, that means predicting a false high detection probability, we apply it to fake synthetic spectra composed of various noise and random fake lines. We use a noise level and a fake line density on a grid of $10 \times 10$ values going from $2.5\times10^{-5}$\,K to $2.5$\,K, and $1$ to $10^{3}$ lines per GHz, both axis following a logarithmic scale. For each grid point, we create 1000 spectra with randomly distributed lines, line-width and intensity range, as in Sect.\,\ref{sub:composite_spectra}. We show in Fig.\,\ref{fig:hallu} the resulting mean detection probabilities for each grid point given by the CNN-model. 

The first result to note is that the detection probabilities are all below 10\% for the parameter space constrained during the training. We also find that the noise level does not impact the detection probabilities, however we observe an increase in this value with increasing fake line density. Still, the detection probability remains bellow 50\% even for the spectra with the maximum line density of 10$^3$. Although the fake lines are randomly distributed, we may expect that a significant number of channels with signal could mimic real emission of some species. Our results suggest that the CNN-model takes into account other information than the position of lines, such as relative intensities. We can therefore conclude that false detections from our CNN-model are unlikely to affect the reliability of the presented results, as long as the line density of the spectra remains in the range of our training sample.

\subsection{Features importance} \label{sub:features}

To analyze the molecular content of millimeter spectra the modeled spectrum of each identified molecule is compared to the observed one. A simultaneous line-fitting is performed in an iterative process to firmly detect emission from a COM, especially from those with a large number of rotational transitions (such as \ce{(CH2OH)2}, \ce{C3H7CN} or \ce{CH3COCH3} from our list of targeted species). In contrast to that, it is more difficult to understand how the CNN-model exactly computes its predictions since it is highly parametric. We aim here to explore the decision making process of our CNN-model by hiding various fractions of spectral features and by doing an occlusion analysis.

\begin{figure}
    \centering
    \includegraphics[width=1\linewidth]{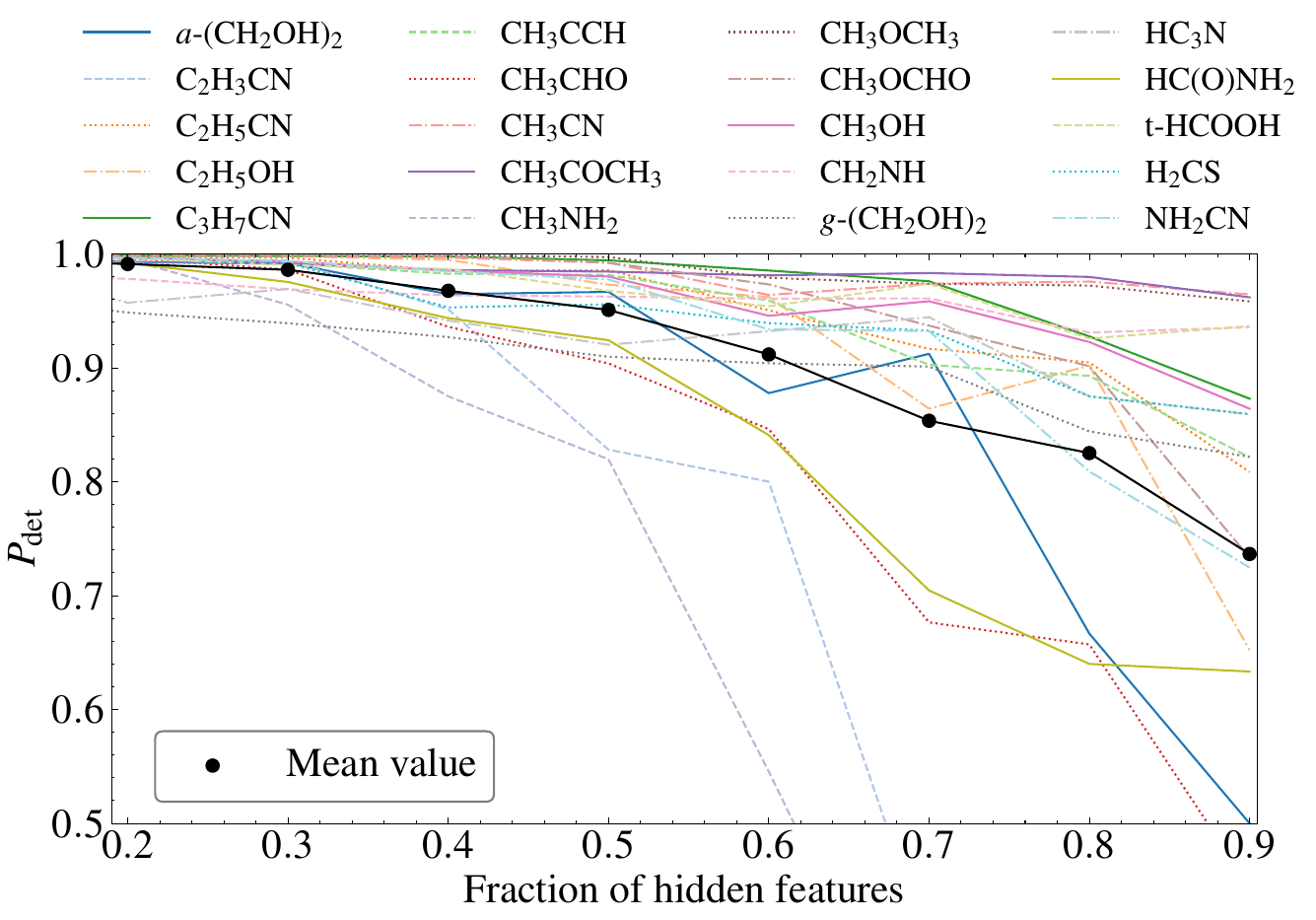}
    \caption{Probability of detection (average on 100 realizations) for each molecule depending on the fraction of intentionally hidden transitions.}
    \label{fig:features_importance}
\end{figure}

We aim to evaluate how the number of transitions impacts the predicted probability for each species individually. We use the spectrum from Sect\,\ref{sub:composite_spectra} (cf. Table\,\ref{tab:parameters_classical_hot_core} with a 5 \kms\ line width and a 50\,mK noise level), and randomly mask an increasing fraction of transitions from 20 to 90\% by steps of 10\% for each molecule separately. We perform $100$ realizations with a different random set of masked lines each time and take the mean model score per molecule to obtain detection probabilities. This allows us to mitigate the fact different lines may have different importance in the detection of a molecule due to brightness and line blending.
The detection probability as a function of the fraction of hidden transitions is shown in Fig.\,\ref{fig:features_importance} for each molecule. All species have a detection probability above $80$\% for less than $50$\,\% of their lines masked, however, the detection probabilities for all species show a decreasing trend with an increasing fraction of masked lines.

Above a 50\% fraction of masked lines, the detection probability for \ce{CH3NH2} and \ce{C2H3CN} drops. Interestingly, \ce{HC3N} still has high detection probability, even when few or no transitions are present. Concerning the other species, the detection probability slightly decreases with the increasing fraction but still remains above $50$\,\%. 
Overall, this result allows us to evaluate the resiliency of the CNN-model to the degradation of the spectral features, but also the level of information redundancy that it learned from the spectra. Our results suggest that this is quite high as we can mask a significant fraction of the lines while still being able to predict the presence of actual molecules with a high probability. Extreme cases, when masking too many transitions of the spectra, become unrealistic and probably correspond to unconstrained cases which would explain some misinterpretations of the model.

However, randomly hiding a fraction of transitions does not allow us to estimate the importance of individual spectral features in the detection process. For this purpose, we perform an occlusion analysis that can help testing the impact of each line on the model prediction. This analysis consists in masking a certain fraction of the signal, referred to as occlusion, and then evaluating the influence of this hidden signal on the CNN-model's classification. The method is illustrated schematically in Sect.\,\ref{appendix:occlusion_analysis} and  Fig.\,\ref{fig:principle_occlusion_analysis}.

We perform this analysis using a sliding window of five channels, where we replace the real signal by the thermal noise of the spectra. We use a step size of three channels for the sliding window, and at each window position compute the difference between the original model score and the one obtained with occlusion. We thus obtain an "occlusion score" at the window position. This occlusion analysis is done for all species individually.

\begin{figure}
    \centering
    \includegraphics[width=1\linewidth]{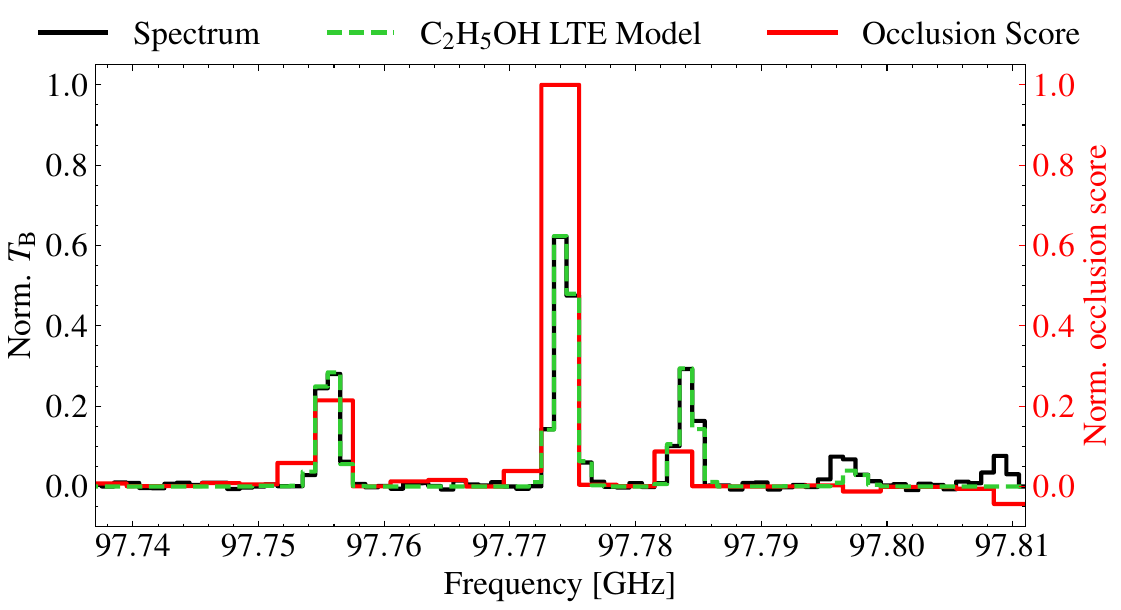}
    \caption{The first maximum of the occlusion analysis score for \ce{C2H5OH}.}
    \label{fig:occlusion}
\end{figure}

The highest the occlusion score, the more a feature is relevant for the detection. We show here the occlusion analysis for \ce{C2H5OH} from the classical hot core synthetic spectrum (cf. Sect.\,\ref{sub:composite_spectra}). A positive occlusion score corresponds to a correlation made by the CNN-model in favor of the molecule to be detected, while a negative score is linked to an anti-correlation. In Fig. \ref{fig:occlusion}, we compare the composite synthetic spectrum, the LTE model of \ce{C2H5OH} and the occlusion score for a small frequency range around the maximum score. This demonstrates that peaks of the occlusion score coincide with transitions from the molecule of interest. Blended lines or lines from other species have a negative score, which reduces the probability of the molecule to be detected. Whereas a transition with a score equals to zero has no impact on the classification of the species.

\subsection{Incomplete frequency coverage of observational setups}\label{sub:setups}

We used for our synthetic spectra a complete spectral coverage between 80.0 and 115.0~GHz. Heterodyne receivers do not cover this frequency range instantaneously, for example the EMIR receiver at the IRAM~30m telescope, provides an instantaneous, non-continuous bandwidth of about $\sim$15.5~GHz \citep{Carter2012}, and can be used to cover our investigated frequency range in two setups. Smaller frequency ranges can also be sufficient to firmly confirm emission from numerous COMs, therefore, we evaluate the performance of our CNN-model on two possible setups from the EMIR receiver in the 3mm band, noting that IRAM NOEMA also provides a very similar instantaneous frequency coverage. Our first setup covers a frequency range of 82--90~GHz and 98--106~GHz (setup 1), while the second setup covers a frequency range of 90--98~GHz, as well as 106--114~GHz (setup 2). Since the CNN-model was trained on spectra covering the full band of $80$ to $115$ GHz, using it on a limited spectral coverage of these setups is not well constrained, implying that the model score can no longer be calibrated or interpreted. To remedy this, it is necessary to readjust its task by implementing transfer learning through a complementary training for each setup of interest in order to recover a more efficient model. Using transfer learning instead of retraining the CNN from its initial state reduces training time and enables effective learning from limited data \citep[e.g.][]{Pan2010,Dominguez2018}. Overall, this  demonstrates that our CNN-model can be efficiently adapted to new observational setups with minimal computational cost.

This complementary training uses the CNN-model parameters and  redefines the output layer to avoid biases. For the rest, we use the same CNN architecture and the same hyperparameters. A special data set is produced for the desired setup with the same procedure as in Sect. \ref{section:training_set} and the data target (i.e., labeling as discussed in Sect.\,\ref{sub:labeling}) is adjusted based on the frequency coverage. Transfer learning was made independently for both setups, and their combination, during $10$ iterations (10\% of the number of examples from the original training). Combining the two setups nearly corresponds to the initially investigated full 80--115~GHz band.

To compare the performance of the CNN-model for these setups we use the diagnostic presented in Sect.\,\ref{sub:metrics}. 
As in the previous sections (Sect.\,\ref{sub:features}), we apply the CNN-model to the classical hot core synthetic spectrum having a 50\,mK noise level in the frequency ranges described above, and using MC-dropout. The number of $5\sigma$ transitions, the obtained median detection probability and the MAD for each molecule depending on the setup are presented in Table\,\ref{tab:emir_setups_results}. The results show a similarly high performance for both setups for the species that meet the detection criterion with typical detection probabilities above 84\% for the majority. The \ce{HC3N} molecule has only one detectable transition in setup 1 and consequently it has a detection probability of $50.0$\,\% with a maximum uncertainty of $8.6$\,\%. This means the molecule is potentially present, however the CNN-model does not have much confidence. On the other hand, \ce{HC3N} has two transitions in setup 2 and the CNN-model is able to robustly identify its spectral signature. The detection probability for \ce{CH3CCH} is nearly zero in setup 2 as no transition lies within this frequency coverage. Similarly, the detection probability of \ce{NH2CN} with one $5\sigma$ transition is low, $26.3$\,\%. When looking at the results of the combined setup, we find that the CNN-model identifies well all molecules.

\begin{table*}[]
    \centering
    \caption{Monte Carlo dropout results for the EMIR setups.}
    \begin{tabular}{c||c|c||c|c||c|c}
    \hline \hline
    \multirow{2}{*}{\textbf{Molecule}} & \multicolumn{2}{c||}{\textbf{EMIR Setup 1}} & 
    \multicolumn{2}{c||}{\textbf{EMIR Setup 2}}& 
    \multicolumn{2}{c}{\textbf{EMIR Setup 1 + 2}}\\
    \cline{2-7}
      & 5$\sigma$ lines & $P_{\rm det}$[\%] & 5$\sigma$ lines & $P_{\rm det}$[\%] & 5$\sigma$ lines & $P_{\rm det}$[\%] \\
     \hline
$a$-(CH$_2$OH)$_2$ & 115 & 99.6$^{+0.3}_{-0.3}$ & 185 & 99.9$^{+0.1}_{-0.1}$ & 300 & 99.9$^{+0.1}_{-0.3}$ \\
C$_2$H$_3$CN & 45 & 99.7$^{+0.2}_{-0.2}$ & 42 & 99.5$^{+0.2}_{-0.2}$ & 87 & 99.7$^{+0.2}_{-0.3}$ \\
C$_2$H$_5$CN & 80 & 99.2$^{+0.5}_{-0.5}$ & 81 & 98.6$^{+0.1}_{-2.4}$ & 161 & 99.8$^{+0.1}_{-0.1}$ \\
C$_2$H$_5$OH & 159 & 99.6$^{+0.2}_{-1.2}$ & 111 & 99.0$^{+0.5}_{-1.2}$ & 270 & 99.9$^{+0.1}_{-0.1}$ \\
C$_3$H$_7$CN & 77 & 99.8$^{+0.1}_{-0.1}$ & 89 & 95.2$^{+2.2}_{-3.6}$ & 166 & 99.9$^{+0.1}_{-0.1}$ \\
CH$_3$CCH & 30 & 98.1$^{+0.7}_{-1.2}$ & 0 & 0.6$^{+0.5}_{-0.2}$ & 30 & 99.7$^{+0.1}_{-0.3}$ \\
CH$_3$CHO & 24 & 99.8$^{+0.1}_{-0.1}$ & 48 & 93.5$^{+1.0}_{-2.5}$ & 72 & 99.8$^{+0.1}_{-0.1}$ \\
CH$_3$CN & 7 & 99.8$^{+0.1}_{-0.1}$ & 21 & 99.8$^{+0.1}_{-0.2}$ & 28 & 99.9$^{+0.1}_{-0.1}$ \\
CH$_3$COCH$_3$ & 15 & 98.0$^{+0.4}_{-1.4}$ & 17 & 86.2$^{+3.7}_{-7.6}$ & 32 & 99.7$^{+0.1}_{-0.1}$ \\
CH$_3$NH$_2$ & 55 & 99.7$^{+0.1}_{-0.2}$ & 42 & 99.4$^{+0.3}_{-0.4}$ & 97 & 99.7$^{+0.1}_{-0.1}$ \\
CH$_3$OCH$_3$ & 43 & 99.6$^{+0.2}_{-0.5}$ & 40 & 99.7$^{+0.1}_{-0.3}$ & 83 & 99.9$^{+0.1}_{-0.1}$ \\
CH$_3$OCHO & 86 & 98.9$^{+0.8}_{-0.4}$ & 75 & 99.6$^{+0.2}_{-0.4}$ & 161 & 99.8$^{+0.1}_{-0.1}$ \\
CH$_3$OH & 55 & 99.5$^{+0.2}_{-0.2}$ & 66 & 99.8$^{+0.1}_{-0.1}$ & 121 & 99.8$^{+0.1}_{-0.1}$ \\
CH$_2$NH & 5 & 84.6$^{+3.1}_{-2.2}$ & 5 & 98.1$^{+0.1}_{-0.1}$ & 10 & 98.1$^{+0.1}_{-0.1}$ \\
$g$-(CH$_2$OH)$_2$ & 27 & 98.7$^{+0.7}_{-0.6}$ & 45 & 93.2$^{+1.7}_{-2.7}$ & 72 & 98.6$^{+0.4}_{-0.7}$ \\
HC$_3$N & 1 & 50.0$^{+8.5}_{-8.6}$ & 2 & 96.5$^{+1.3}_{-1.1}$ & 3 & 99.1$^{+0.3}_{-0.1}$ \\
HC(O)NH$_2$ & 35 & 99.6$^{+0.1}_{-0.4}$ & 15 & 99.0$^{+0.5}_{-0.7}$ & 50 & 99.7$^{+0.1}_{-0.1}$ \\
t-HCOOH & 9 & 95.0$^{+1.0}_{-1.4}$ & 13 & 98.5$^{+0.5}_{-2.3}$ & 22 & 99.7$^{+0.1}_{-0.1}$ \\
H$_2$CS & 10 & 99.5$^{+0.1}_{-0.1}$ & 3 & 86.2$^{+3.2}_{-5.8}$ & 13 & 99.5$^{+0.1}_{-0.1}$ \\
NH$_2$CN & 13 & 98.8$^{+0.3}_{-0.9}$ & 1 & 26.3$^{+7.0}_{-5.8}$ & 14 & 99.7$^{+0.1}_{-0.1}$ \\

   \hline
    \end{tabular}
    \label{tab:emir_setups_results}
\end{table*}

\section{Application to observational data} \label{section:obs_application}

\subsection{Predictions using archival wide-band observational data}
As shown in Sect.\,\ref{sub:line_density} and \ref{sub:noise}, our full 80--115\,GHz band CNN-model has a well-defined parameter space for the observed noise and line density. Sect.\,\ref{sub:performance} and \ref{sub:false_positives} describe its performance on synthetic spectra and suggest an overall high and reliable performance. Here we aim to apply this model to observational data, that is a significant step forward in testing the applicability of our CNN-model. In contrast to the synthetic spectra, observational data may contain emission from species not considered in our synthetic composite spectra and abundance ratios outside of that of our training set. Furthermore, gas kinematics above the 1~MHz resolution is neglected in our simulated data, which may hinder the detection performance of the CNN-model.

\begin{table*}[]
    \caption{Observational data.}
    \centering
    \begin{tabular}{lccccccl}
    \hline \hline Source & Telescope &  Freq. coverage & Resolution & Noise & Line density & References\\
    & & (GHz) & (kHz) & (mK) & & \\
    \hline
Sgr B2(N) & IRAM 30m & 79.989 -- 115.984 & 320 & 24  & 102  &\citet{Belloche2013} \\
Sgr B2(M) & IRAM 30m & 79.989 -- 115.984 & 320 & 25  & 26  & \citet{Belloche2013} \\
G34.26+0.15 & IRAM 30m & 84.276 -- 115.745 & 200 & 20 (9.4) & (19.8) & \citet{Csengeri2016B} \\
CygX-N63 & IRAM 30m & 82.350 -- 117.502 & 200 & 2.9 (1.9) & (12.7) & \citet{Fechtenbaum2015} \\

   \hline
    \end{tabular}
    \tablefoot{In parenthesis the measured noise and computed line density for the resampled spectra to a resolution of 1\,MHz.} 
    \label{tab:obslist}
\end{table*}

We use observations obtained with the EMIR receiver at the IRAM 30m telescope toward chemically rich regions with hot-core characteristics from the literature that have a similar total frequency coverage as in our models and where a proper modeling of the spectra has already been performed. We investigate the archetypal hot cores Sgr~B2(N) and B2(M) \citep{Belloche2013}, G34.26+0.15 \citep{Csengeri2016B}, and the pre-hot core CygX-N63 \citep{fecht2015}. Since the CNN-model is expected to be able to work on any continuous spectrum having the same frequency coverage and resolution as used for network training, it is expected to be applicable to interferometric data as well.

Basic data reduction steps have already been applied to the data, such as baseline subtraction and correcting the frequency axis for the source \vlsr. The observational data have a spectral resolution better than 1~MHz and where thus first resampled to the same frequency axis and resolution as that of the training set. Preparation of input spectra to the CNN-model is described in more detail in App.\,\ref{appendix:application}. The noise is measured in both the original and the resampled spectra, and we measure the line density  as in Sect.\,\ref{sub:line_density}. We use the line density information to obtain an initial characterization of the spectra (see also Sect.\,\ref{sub:line_density}) and to situate the observational data within the parameter space of the training set. As discussed in \citet{Belloche2013}, the spectra of the Sgr~B2 sources do not reach the confusion limit. Since the rest of the here studies sources have lower line-widths and line densities, they are also above the confusion limit. The corresponding parameters of the used observational data are listed in Table~\ref{tab:obslist}.

We assign a detection status for each source and species partly based on literature results and by re-analyzing the data as resampling may dilute the intensity of spectrally unresolved signal and increase the impact of line blending. For this, we visually inspect the resampled spectra and compare it with synthetic LTE models, however, without performing a proper fitting because our results suggest little change in the overall detection of molecules toward these spectra. We list these results in Table\,\ref{tab:pred_mm_data} as either firm observational detections from the literature or as tentative detections, where a more precise modeling would be needed to confirm the detectability of these species.

Our CNN-model is used to predict the molecular composition of the observational data. As in Sect.\,\ref{sub:noise}, we apply the CNN-model with 100 realizations of MC-dropout to obtain a median detection probability for each species listed in Table\,\ref{tab:pred_mm_data}. We compare the detection probability of the CNN-model for each molecule, and discuss these values in the context of the observational results.

\subsubsection{Sgr~B2(N)}

We use the publicly available  IRAM~30m observations from \citet{Belloche2013} to investigate the spectrum of Sgr~B2(N). The CNN-model predicts emission from \ce{C2H3CN}, \ce{C2H5CN}, \ce{C2H5OH}, \ce{CH3CCH}, \ce{CH3CN}, \ce{CH3OCH3}, \ce{CH3OCHO}, \ce{CH3OH}, \ce{HC3N}, \ce{HC(O)NH2}, \ce{H2CS} and \ce{NH2CN} with a high detection probability of $P_{\rm det}>66$\%. On the other hand, it finds a very low detection probability for species that have weak, nevertheless observationally detectable signatures in the spectrum, such as $a$-\ce{(CH2OH)2}, \ce{C3H7CN}, \ce{CH3NH2}, \ce{CH2NH} which means that their transitions are difficult to identify. Considering our criteria of detectability we would have expected the model to predict higher probabilities for these species. In addition, it gives a detection probability of 38.7\% for \ce{CH3CHO}, although with uncertainties of $\sim$9.2--12.9\%, and 32.3\% with similarly large uncertainties for \ce{CH3COCH3}. From these examples \ce{CH3CHO} is expected to be a relatively easily detectable species. $g$-\ce{(CH2OH)2} is predicted with a probability of 21.7\,\%, with an error of $\sim$7--12\% even though its spectral signature is observationally not confirmed. As shown in Sect\,\ref{sub:false_positives}, we find this kind of behavior when the CNN-model is applied to a parameter range, or conditions it has not been trained for.

\subsubsection{Sgr~B2(M)}
For Sgr~B2(M), the results of the CNN-model seem globally less convincing than for Sgr~B2(N). 
The detection probability is generally lower for heavier COMs compared to Sgr~B2(N), but high values ($>$80\%) are found for more abundant species like \ce{CH3CCH}, \ce{CH3CN}, \ce{CH3OH} and \ce{HC3N}. We find detection probabilities in the range of 35 to 50\% for species, such as \ce{CH3CHO}, \ce{CH3OCHO}, \ce{HC(O)NH2}, \ce{H2CS} and \ce{NH2CN}. The error-bars on these predictions are larger of $\sim$5.6--14.3, and imply that the CNN-model cannot properly predict the presence of these species toward this source. Nevertheless, it correctly predicts very low probabilities of $<$3\% for non-detected species, such as \ce{t-HCOOH}, $a$-\ce{(CH2OH)2} and $g$-\ce{(CH2OH)2}. 

This systematically low probability for molecules that should be identified suggests that the spectrum of this source is noticeably different from the training sample. When comparing the spectrum of Sgr~B2(N) and Sgr~B2(M), both have an important number of absorption lines due to foreground clouds \citep[cf.][]{Thiel2019}. However, Sgr~B2(M) exhibits significantly less spectral lines, making its overall characteristics less line rich, and the number of absorption lines compared to the emission lines more prominent compared to Sgr~B2(N), and the training set.

\subsubsection{G34.26+0.15}
We also include in our analysis the archetypal hot core G34.26+0.15, where the spectrum is taken from \citet{Csengeri2016B}, and the molecular composition of this frequency range was already discussed in (\citealp{Mookerjea2007}, \citealp{WidicusWeaver2017} for higher frequency observations). This source shows less complex kinematic features, and no significant absorption features from foreground clouds compared to the ones of Sgr~B2 sources. 
We find that the predicted probability of our CNN-model is high, $\gtrsim$84\% for \ce{C2H5CN}, \ce{CH3CCH}, \ce{CH3CN}, \ce{CH3OCH3}, \ce{CH3OCHO}, \ce{CH3OH}, \ce{HC3N}, \ce{HC(O)NH2}, and \ce{H2CS}. Low detection probability of 2.7-21.6\% is found for \ce{C2H5OH}, \ce{CH3CHO}, \ce{CH2NH} and \ce{t-HCOOH} with expected detections, while non-detections have low probability values.

\subsubsection{CygX-N63}

Here we also investigate the spectrum of CygX-N63 that is a precursor of a hot core \citep{fecht2015}, a source that is less line-rich compared to the other ones with no absorption features. In this respect, this spectrum represents the simplest use case, where we expect reliable results. The CNN-model detects high detection probabilities for \ce{CH3CCH}, \ce{CH3CHO}, \ce{CH3CN}, \ce{CH3OH}, \ce{H3CN}, and \ce{H2CS}. The COMs \ce{CH3OCH3} and \ce{HC(O)NH2} present probabilities slightly above 50\,\%, whereas the others are less than 30\,\% with \ce{CH3OCHO} and \ce{t-HCOOH} having a value around 20\,\%. There are no false high probability detections, consistent with our previous results.

\begin{table*}[]
\caption{Detection probabilities obtained with MC dropout for millimeter data.}
    \centering
    \begin{tabular}{c||c|c||c|c||c|c||c|c}
    \hline \hline
    \multirow{2}{*}{\textbf{Molecule}} & \multicolumn{2}{c||}{\textbf{Sgr~B2(N)}} & 
    \multicolumn{2}{c ||}{\textbf{Sgr~B2(M)}} & \multicolumn{2}{c ||}{\textbf{G34.26+0.15}} & \multicolumn{2}{c }{\textbf{CygX-N63}}  \\
    \cline{2-9}
       &    $P_{\rm det}$[\%]    &  Detection   &   $P_{\rm det}$[\%] &  Detection   &    $P_{\rm det}$[\%] &  Detection   &   $P_{\rm det}$[\%]  & Detection \\
\hline

$a$-(CH$_2$OH)$_2$ & 3.6$^{+1.7}_{-1.3}$ & D & 0.2$^{+0.3}_{-0.1}$ & N & 3.4$^{+1.9}_{-1.6}$ & N & 0.2$^{+0.3}_{-0.1}$ & N  \\
C$_2$H$_3$CN & 96.1$^{+1.6}_{-1.6}$ & D & 8.5$^{+6.3}_{-2.9}$ & D & 22.0$^{+9.5}_{-5.7}$ & T{\bf$^\ddagger$} & 0.3$^{+0.1}_{-0.1}$ & D  \\
C$_2$H$_5$CN & 98.8$^{+0.4}_{-0.2}$ & D & 27.9$^{+8.0}_{-9.6}$ & D & 84.3$^{+5.3}_{-11.4}$ & D & 3.0$^{+2.1}_{-0.9}$ & D  \\
C$_2$H$_5$OH & 74.2$^{+6.9}_{-11.2}$ & D & 0.5$^{+0.5}_{-0.1}$ & D & 13.0$^{+5.1}_{-5.0}$ & D & 3.7$^{+3.0}_{-1.1}$ & D  \\
C$_3$H$_7$CN & 8.1$^{+3.4}_{-2.5}$ & D$^\dagger$ & 1.0$^{+0.4}_{-0.4}$ & N & 0.1$^{+0.1}_{-0.1}$ & N & 0.9$^{+0.5}_{-0.4}$ & N  \\
CH$_3$CCH & 96.3$^{+1.2}_{-2.4}$ & D & 83.7$^{+7.9}_{-9.1}$ & D & 95.7$^{+1.4}_{-2.1}$ & D & 97.1$^{+1.4}_{-1.2}$ & D  \\
CH$_3$CHO & 38.7$^{+12.9}_{-9.2}$ & D & 37.9$^{+10.3}_{-8.5}$ & D & 21.6$^{+8.5}_{-9.1}$ & D & 89.4$^{+4.1}_{-8.3}$ & D \\
CH$_3$CN & 98.6$^{+0.7}_{-1.9}$ & D & 84.2$^{+3.7}_{-4.2}$ & D & 99.9$^{+0.1}_{-0.1}$ & D & 96.9$^{+0.9}_{-2.2}$ & D  \\
CH$_3$COCH$_3$ & 32.3$^{+12.3}_{-9.4}$ & D & 2.0$^{+0.3}_{-0.6}$ & D & 50.0$^{+15.7}_{-12.5}$ & T{\bf$^\ddagger$} & 6.5$^{+4.6}_{-1.8}$ & D  \\
CH$_3$NH$_2$ & 2.2$^{+0.5}_{-0.2}$ & D & 4.3$^{+1.7}_{-1.7}$ & D & 2.4$^{+0.6}_{-0.4}$ & T & 4.3$^{+2.0}_{-1.6}$ & N  \\
CH$_3$OCH$_3$ & 65.8$^{+7.5}_{-8.8}$ & D & 0.3$^{+0.2}_{-0.1}$ & D & 84.4$^{+6.2}_{-4.4}$ & D & 57.7$^{+10.9}_{-16.0}$ & D  \\
CH$_3$OCHO & 90.0$^{+3.8}_{-2.8}$ & D & 42.9$^{+11.7}_{-14.3}$ & D & 98.3$^{+0.5}_{-1.6}$ & D & 29.8$^{+11.9}_{-13.6}$ & D  \\
CH$_3$OH & 99.8$^{+0.1}_{-0.1}$ & D & 98.2$^{+1.0}_{-0.6}$ & D & 99.3$^{+0.2}_{-0.6}$ & D & 99.8$^{+0.1}_{-0.1}$ & D  \\
CH$_2$NH & 19.9$^{+4.3}_{-2.9}$ & D & 5.2$^{+2.5}_{-1.1}$ & D & 2.7$^{+0.7}_{-0.7}$ & D & 0.4$^{+0.1}_{-0.1}$ & N  \\
$g$-(CH$_2$OH)$_2$ & 21.7$^{+12.0}_{-7.1}$ & N & 0.3$^{+0.1}_{-0.1}$ & N & 0.3$^{+0.4}_{-0.1}$ & N & 2.8$^{+2.2}_{-1.8}$ & N  \\
HC$_3$N & 99.3$^{+0.1}_{-0.2}$ & D & 99.6$^{+0.1}_{-0.1}$ & D & 99.3$^{+0.2}_{-0.2}$ & D & 99.1$^{+0.2}_{-0.1}$ & D  \\
HC(O)NH$_2$ & 79.6$^{+6.5}_{-8.1}$ & D & 43.4$^{+13.5}_{-11.2}$ & D & 90.3$^{+2.3}_{-7.6}$ & D & 56.1$^{+11.4}_{-9.9}$ & D\\
t-HCOOH & 42.8$^{+9.7}_{-11.4}$ & D & 2.6$^{+0.9}_{-0.4}$ & N & 13.7$^{+3.2}_{-5.2}$ & D & 18.5$^{+8.9}_{-3.3}$ & D   \\
H$_2$CS & 98.8$^{+0.4}_{-1.0}$ & D & 41.0$^{+5.6}_{-11.2}$ & D & 98.9$^{+0.3}_{-0.8}$ & D & 99.3$^{+0.1}_{-0.3}$ & D  \\
NH$_2$CN & 86.0$^{+6.6}_{-3.5}$ & D & 45.1$^{+7.4}_{-10.0}$ & D & 1.4$^{+1.2}_{-0.2}$ & N & 0.7$^{+0.4}_{-0.3}$ & N  \\
\hline

    \end{tabular}
    \tablefoot{The detection status for each species corresponds to detection (D), non detection (N), or tentative detection (T) in the corresponding spectrum. $^\dagger$ Only the normal form was detected in this spectrum.  $^\ddagger$: Although at higher frequencies \citet{WidicusWeaver2017} detects emission from this species.}
    
    \label{tab:pred_mm_data}
\end{table*}

\subsection{Applicability and limitation of the method} \label{sub:discussion_observations}
 
We successfully generalized our CNN-model trained on synthesized data to observational data. Overall, we obtain large probabilities ($>50$\%) for most of the abundant species with expected detections, and we do not find striking false positives for the investigated spectra, which is a very encouraging result. While the detection probabilities were typically higher ($>90$\%) for synthetic spectra on a well-constrained parameter space, we find lower detection probabilities corresponding to a range of 20-80\% for observational data, which is likely due to the differences between real observations and our synthetic spectra.
Low detection probabilities found for some COMs may be due to their signal being weaker than $5\sigma$, and also due to significant line blending when resampling the spectra to 1~MHz resolution that is particularly limiting the application to sources like Sgr~B2(N). We notice a systematically lower performance of our CNN-model for \ce{CH3NH2} and \ce{CH2NH} which may come from blended lines as discussed in Sect.\,\ref{sub:performance}.

The CNN-model presented in this paper was designed to detect molecules in rich millimeter spectra. The obtained results demonstrate  that the CNN-model is able to retrieve the spectroscopic fingerprint of species, although when applied to observational data, the most reliable results are obtained for the well known most abundant species. We do not have evidence for a non-uniform noise over the band to cause a major issue, neither the line kinematics that is present in the observational data at $>$1~MHz resolution, such as the strong absorption features for the Sgr~B2 sources, or high-velocity emission from outflows.
Neither can we identify any systematic trend in the performance that could be specific to a species.

The current version of the CNN-model reaches its limits when applying to extreme sources like for Sgr~B2(N), a limitation that mostly comes from a large number of lines originating from species not included in our LTE modeling, and the spectral resolution that implies significant line blending. For the rest of the tested sources, such as G34.26+0.15 and CygX-N63, line blending is less of an issue, however, the here used spectral resolution is insufficient to resolve individual lines.

\section{Conclusions}
This study aimed to validate the capacity of artificial neural networks to learn the spectroscopic signatures of various molecules using synthetic LTE model spectra.
The main challenge was to handle the large parameter space of the training set and to optimize the architecture for the best performance. Our primary result is that the CNN-model provides robust and reliable results on synthetic data, demonstrating that it is able to learn signatures of  rotational spectra specific to molecular species.
The overall high performance indicates a meaningful application for identifying the molecular content even in line-rich spectra. 

We optimized the CNN-model to be able to treat 20 species and a spectral resolution of 1\,MHz for a frequency coverage of 80-115~GHz. Our analysis revealed that the CNN-model performs well for a wide range of line density and noise. We also show that the CNN-model takes into account not only the position of lines, but information from the entire spectrum to perform its predictions. A notable strength of our CNN-model is its robustness against false positives.

We demonstrated that it is possible to adapt our CNN-model to different frequency coverages corresponding to observational frequency ranges provided by the IRAM 30m and NOEMA telescopes through transfer learning. Application to observations is demonstrated with meaningful detection probabilities confirming that CNN-models are a promising approach to extract information on the molecular composition of millimeter spectra. A better frequency sampling would increase the applicability of our model to even higher line density regimes, and better treat line blending. Further improvements require the development of more powerful CNN architectures, and a training dataset including more molecular species. While this work focuses on methodological development and validation, our approach opens a promising potential for statistical applications on large datasets, allowing us to explore connections between molecular composition and source type, evolutionary stage, or galactic environment.

\bibliographystyle{aa} 
\bibliography{kessler.bib}

\begin{thebibliography}{95}
\expandafter\ifx\csname natexlab\endcsname\relax\def\natexlab#1{#1}\fi

\bibitem[{Araujo {et~al.}(2019)Araujo, Norris, \& Sim}]{Araujo2019}
Araujo, A., Norris, W., \& Sim, J. 2019, Distill,
  https://distill.pub/2019/computing-receptive-fields

\bibitem[{{Baek} {et~al.}(2022){Baek}, {Lee}, {Hirota}, {Kim}, \&
  {Kim}}]{Baek2022}
{Baek}, G., {Lee}, J.-E., {Hirota}, T., {Kim}, K.-T., \& {Kim}, M.~K. 2022,
  \apj, 939, 84

\bibitem[{{Bailer-Jones} {et~al.}(1997){Bailer-Jones}, {Irwin}, {Gilmore}, \&
  {von Hippel}}]{Bailer-Jones1997}
{Bailer-Jones}, C. A.~L., {Irwin}, M., {Gilmore}, G., \& {von Hippel}, T. 1997,
  \mnras, 292, 157

\bibitem[{{Behrens} {et~al.}(2024){Behrens}, {Mangum}, {Viti}, {Holdship},
  {Huang}, {Bouvier}, {Butterworth}, {Eibensteiner}, {Harada}, {Mart{\'\i}n},
  {Sakamoto}, {Muller}, {Tanaka}, {Colzi}, {Henkel}, {Meier}, {Rivilla}, \&
  {van der Werf}}]{Behrens2024}
{Behrens}, E., {Mangum}, J.~G., {Viti}, S., {et~al.} 2024, \apj, 977, 38

\bibitem[{{Belloche} {et~al.}(2025){Belloche}, {Garrod}, {M{\"u}ller}, {Morin},
  {Willis}, \& {Menten}}]{Belloche2025}
{Belloche}, A., {Garrod}, R.~T., {M{\"u}ller}, H.~S.~P., {et~al.} 2025, \aap,
  698, A143

\bibitem[{{Belloche} {et~al.}(2022){Belloche}, {Garrod}, {Zingsheim},
  {M{\"u}ller}, \& {Menten}}]{Belloche2022}
{Belloche}, A., {Garrod}, R.~T., {Zingsheim}, O., {M{\"u}ller}, H.~S.~P., \&
  {Menten}, K.~M. 2022, \aap, 662, A110

\bibitem[{{Belloche} {et~al.}(2020){Belloche}, {Maury}, {Maret}, {Anderl},
  {Bacmann}, {Andr{\'e}}, {Bontemps}, {Cabrit}, {Codella}, {Gaudel}, {Gueth},
  {Lef{\`e}vre}, {Lefloch}, {Podio}, \& {Testi}}]{Belloche2020}
{Belloche}, A., {Maury}, A.~J., {Maret}, S., {et~al.} 2020, \aap, 635, A198

\bibitem[{{Belloche} {et~al.}(2013){Belloche}, {M{\"u}ller}, {Menten},
  {Schilke}, \& {Comito}}]{Belloche2013}
{Belloche}, A., {M{\"u}ller}, H.~S.~P., {Menten}, K.~M., {Schilke}, P., \&
  {Comito}, C. 2013, \aap, 559, A47

\bibitem[{Bengio \& Glorot(2010)}]{Bengio2010}
Bengio, Y. \& Glorot, X. 2010, International Conference on Artificial
  Intelligence and Statistics, 249

\bibitem[{{Bonfand} {et~al.}(2017){Bonfand}, {Belloche}, {Menten}, {Garrod}, \&
  {M{\"u}ller}}]{Bonfand2017}
{Bonfand}, M., {Belloche}, A., {Menten}, K.~M., {Garrod}, R.~T., \&
  {M{\"u}ller}, H.~S.~P. 2017, \aap, 604, A60

\bibitem[{{Bouscasse} {et~al.}(2022){Bouscasse}, {Csengeri}, {Belloche},
  {Wyrowski}, {Bontemps}, {G{\"u}sten}, \& {Menten}}]{Bouscasse2022}
{Bouscasse}, L., {Csengeri}, T., {Belloche}, A., {et~al.} 2022, \aap, 662, A32

\bibitem[{{Bouscasse} {et~al.}(2024){Bouscasse}, {Csengeri}, {Wyrowski},
  {Menten}, \& {Bontemps}}]{Bouscasse2024}
{Bouscasse}, L., {Csengeri}, T., {Wyrowski}, F., {Menten}, K.~M., \&
  {Bontemps}, S. 2024, \aap, 686, A252

\bibitem[{{Bouvier} {et~al.}(2025){Bouvier}, {Viti}, {Mangum}, {Eibensteiner},
  {Behrens}, {Rivilla}, {L{\'o}pez-Gallifa}, {Mart{\'\i}n}, {Harada}, {Muller},
  {Colzi}, \& {Sakamoto}}]{Bouvier2025}
{Bouvier}, M., {Viti}, S., {Mangum}, J.~G., {et~al.} 2025, \aap, 698, A261

\bibitem[{Bradley(1997)}]{Bradley1997}
Bradley, A.~P. 1997, Pattern Recognition, 30, 1145

\bibitem[{{Carter, M.} {et~al.}(2012){Carter, M.}, {Lazareff, B.}, {Maier, D.},
  {Chenu, J.-Y.}, {Fontana, A.-L.}, {Bortolotti, Y.}, {Boucher, C.},
  {Navarrini, A.}, {Blanchet, S.}, {Greve, A.}, {John, D.}, {Kramer, C.},
  {Morel, F.}, {Navarro, S.}, {Peñalver, J.}, {Schuster, K. F.}, \& {Thum,
  C.}}]{Carter2012}
{Carter, M.}, {Lazareff, B.}, {Maier, D.}, {et~al.} 2012, A\&A, 538, A89

\bibitem[{{CASA Team} {et~al.}(2022){CASA Team}, {Bean}, {Bhatnagar}, {Castro},
  {Donovan Meyer}, {Emonts}, {Garcia}, {Garwood}, {Golap}, {Gonzalez Villalba},
  {Harris}, {Hayashi}, {Hoskins}, {Hsieh}, {Jagannathan}, {Kawasaki},
  {Keimpema}, {Kettenis}, {Lopez}, {Marvil}, {Masters}, {McNichols},
  {Mehringer}, {Miel}, {Moellenbrock}, {Montesino}, {Nakazato}, {Ott}, {Petry},
  {Pokorny}, {Raba}, {Rau}, {Schiebel}, {Schweighart}, {Sekhar}, {Shimada},
  {Small}, {Steeb}, {Sugimoto}, {Suoranta}, {Tsutsumi}, {van Bemmel},
  {Verkouter}, {Wells}, {Xiong}, {Szomoru}, {Griffith}, {Glendenning}, \&
  {Kern}}]{CASATeam2022}
{CASA Team}, {Bean}, B., {Bhatnagar}, S., {et~al.} 2022, \pasp, 134, 114501

\bibitem[{{Caselli} \& {Ceccarelli}(2012)}]{Caselli2012}
{Caselli}, P. \& {Ceccarelli}, C. 2012, \aapr, 20, 56

\bibitem[{{Caux} {et~al.}(2011){Caux}, {Bottinelli}, {Vastel}, \&
  {Glorian}}]{Caux2011}
{Caux}, E., {Bottinelli}, S., {Vastel}, C., \& {Glorian}, J.~M. 2011, in IAU
  Symposium, Vol. 280, The Molecular Universe, ed. J.~{Cernicharo} \&
  R.~{Bachiller}, 120

\bibitem[{{Ceccarelli} {et~al.}(2022){Ceccarelli}, {Codella}, {Balucani},
  {Bockel{\'e}e-Morvan}, {Herbst}, {Vastel}, {Caselli}, {Favre}, {Lefloch}, \&
  {{\"O}berg}}]{Ceccarelli2022}
{Ceccarelli}, C., {Codella}, C., {Balucani}, N., {et~al.} 2022, arXiv e-prints,
  arXiv:2206.13270

\bibitem[{{Chin} {et~al.}(1996){Chin}, {Henkel}, {Whiteoak}, {Langer}, \&
  {Churchwell}}]{Chin1996}
{Chin}, Y.~N., {Henkel}, C., {Whiteoak}, J.~B., {Langer}, N., \& {Churchwell},
  E.~B. 1996, \aap, 305, 960

\bibitem[{{Coletta} {et~al.}(2020){Coletta}, {Fontani}, {Rivilla}, {Mininni},
  {Colzi}, {S{\'a}nchez-Monge}, \& {Beltr{\'a}n}}]{Coletta2020}
{Coletta}, A., {Fontani}, F., {Rivilla}, V.~M., {et~al.} 2020, \aap, 641, A54

\bibitem[{{Cornu}(2025)}]{Cornu2025}
{Cornu}, D. 2025, {CIANNA: Convolutional Interactive Artificial Neural Networks
  by/for Astrophysicists}, Astrophysics Source Code Library, record
  ascl:2501.005

\bibitem[{{Csengeri} {et~al.}(2016){Csengeri}, {Leurini}, {Wyrowski},
  {Urquhart}, {Menten}, {Walmsley}, {Bontemps}, {Wienen}, {Beuther}, {Motte},
  {Nguyen-Luong}, {Schilke}, {Schuller}, {Zavagno}, \& {Sanna}}]{Csengeri2016B}
{Csengeri}, T., {Leurini}, S., {Wyrowski}, F., {et~al.} 2016, \aap, 586, A149

\bibitem[{{Dom\'inguez Sánchez} {et~al.}(2018){Dom\'inguez Sánchez},
  {Huertas-Company}, {Bernardi}, {Kaviraj}, {Fischer}, {Abbott}, {Abdalla},
  {Annis}, {Avila}, {Brooks}, {Buckley-Geer}, {Carnero Rosell}, {Carrasco
  Kind}, {Carretero}, {Cunha}, {D’Andrea}, {da Costa}, {Davis}, {De Vicente},
  {Doel}, {Evrard}, {Fosalba}, {Frieman}, {Garc\'ia-Bellido}, {Gaztanaga},
  {Gerdes}, {Gruen}, {Gruendl}, {Gschwend}, {Gutierrez}, {Hartley},
  {Hollowood}, {Honscheid}, {Hoyle}, {James}, {Kuehn}, {Kuropatkin}, {Lahav},
  {Maia}, {March}, {Melchior}, {Menanteau}, {Miquel}, {Nord}, {Plazas},
  {Sanchez}, {Scarpine}, {Schindler}, {Schubnell}, {Smith}, {Smith},
  {Soares-Santos}, {Sobreira}, {Suchyta}, {Swanson}, {Tarle}, {Thomas},
  {Walker}, \& {Zuntz}}]{Dominguez2018}
{Dom\'inguez Sánchez}, H., {Huertas-Company}, M., {Bernardi}, M., {et~al.}
  2018, Monthly Notices of the Royal Astronomical Society, 484, 93

\bibitem[{{Drozdovskaya} {et~al.}(2019){Drozdovskaya}, {van Dishoeck}, {Rubin},
  {J{\o}rgensen}, \& {Altwegg}}]{Drozdovskaya2019}
{Drozdovskaya}, M.~N., {van Dishoeck}, E.~F., {Rubin}, M., {J{\o}rgensen},
  J.~K., \& {Altwegg}, K. 2019, \mnras, 490, 50

\bibitem[{{Dupourqu{\'e}} {et~al.}(2024){Dupourqu{\'e}}, {Clerc},
  {Pointecouteau}, {Eckert}, {Gaspari}, {Lovisari}, {Pratt}, {Rasia},
  {Rossetti}, {Vazza}, {Balboni}, {Bartalucci}, {Bourdin}, {De Luca}, {De
  Petris}, {Ettori}, {Ghizzardi}, \& {Mazzotta}}]{Dupourque2024}
{Dupourqu{\'e}}, S., {Clerc}, N., {Pointecouteau}, E., {et~al.} 2024, \aap,
  687, A58

\bibitem[{{Duronea} {et~al.}(2019){Duronea}, {Bronfman}, {Mendoza}, {Merello},
  {Finger}, {Reyes}, {Herv{\'\i}as-Caimapo}, {Faure}, {Cappa}, {Arnal},
  {L{\'e}pine}, {Kleiner}, \& {Nyman}}]{Duronea2019}
{Duronea}, N.~U., {Bronfman}, L., {Mendoza}, E., {et~al.} 2019, \mnras, 489,
  1519

\bibitem[{{Einig} {et~al.}(2024){Einig}, {Palud}, {Roueff}, {Pety}, {Bron}, {Le
  Petit}, {Gerin}, {Chanussot}, {Chainais}, {Thouvenin}, {Languignon},
  {Be{\v{s}}li{\'c}}, {Coud{\'e}}, {Mazurek}, {Orkisz}, {Santa-Maria},
  {S{\'e}gal}, {Zakardjian}, {Bardeau}, {Demyk}, {de Souza Magalh{\~a}es},
  {Goicoechea}, {Gratier}, {Guzm{\'a}n}, {Hughes}, {Levrier}, {Le Bourlot},
  {Lis}, {Liszt}, {Peretto}, {Roueff}, \& {Sievers}}]{Einig2024}
{Einig}, L., {Palud}, P., {Roueff}, A., {et~al.} 2024, \aap, 691, A109

\bibitem[{{Endres} {et~al.}(2016){Endres}, {Schlemmer}, {Schilke}, {Stutzki},
  \& {M{\"u}ller}}]{Endres2016}
{Endres}, C.~P., {Schlemmer}, S., {Schilke}, P., {Stutzki}, J., \&
  {M{\"u}ller}, H. S.~P. 2016, Journal of Molecular Spectroscopy, 327, 95

\bibitem[{{Fechtenbaum}(2015)}]{fecht2015}
{Fechtenbaum}, S. 2015, PhD thesis, Université de Bordeaux, thèse de doctorat
  dirigée par Bontemps, Sylvain Astrophysique, plasmas, nucléaire Bordeaux
  2015

\bibitem[{{Fechtenbaum} {et~al.}(2015){Fechtenbaum}, {Bontemps}, {Schneider},
  {Csengeri}, {Duarte-Cabral}, {Herpin}, \& {Lefloch}}]{Fechtenbaum2015}
{Fechtenbaum}, S., {Bontemps}, S., {Schneider}, N., {et~al.} 2015, \aap, 574,
  L4

\bibitem[{{Feng} {et~al.}(2015){Feng}, {Beuther}, {Henning}, {Semenov},
  {Palau}, \& {Mills}}]{Feng2015}
{Feng}, S., {Beuther}, H., {Henning}, T., {et~al.} 2015, \aap, 581, A71

\bibitem[{{Fried} {et~al.}(2023){Fried}, {Lee}, {Byrne}, \&
  {McGuire}}]{Fried2023}
{Fried}, Z. T.~P., {Lee}, K. L.~K., {Byrne}, A.~N., \& {McGuire}, B.~A. 2023,
  arXiv e-prints, arXiv:2305.11193

\bibitem[{{Fried} \& {McGuire}(2024)}]{Fried2024}
{Fried}, Z. T.~P. \& {McGuire}, B.~A. 2024, arXiv e-prints, arXiv:2408.15819

\bibitem[{{Gal} \& {Ghahramani}(2015)}]{Gal2015}
{Gal}, Y. \& {Ghahramani}, Z. 2015, arXiv e-prints, arXiv:1506.02142

\bibitem[{{Garrod} \& {Herbst}(2006)}]{Garrod2006}
{Garrod}, R.~T. \& {Herbst}, E. 2006, \aap, 457, 927

\bibitem[{{Garrod} {et~al.}(2022){Garrod}, {Jin}, {Matis}, {Jones}, {Willis},
  \& {Herbst}}]{Garrod2022}
{Garrod}, R.~T., {Jin}, M., {Matis}, K.~A., {et~al.} 2022, \apjs, 259, 1

\bibitem[{{Giannetti} {et~al.}(2025){Giannetti}, {Leurini}, {Schisano},
  {Casasola}, {Pillai}, {Sanna}, \& {Ferrada-Chamorro}}]{Giannetti2025}
{Giannetti}, A., {Leurini}, S., {Schisano}, E., {et~al.} 2025, \aap, 698, A90

\bibitem[{{Giese} {et~al.}(2024){Giese}, {Thompson}, {Lis}, \& {Widicus
  Weaver}}]{Giese2024}
{Giese}, M.~M., {Thompson}, W.~E., {Lis}, D.~C., \& {Widicus Weaver}, S.~L.
  2024, \apj, 960, 6

\bibitem[{{Ginsburg} \& {Mirocha}(2011)}]{Ginsburg2011}
{Ginsburg}, A. \& {Mirocha}, J. 2011, {PySpecKit: Python Spectroscopic
  Toolkit}, Astrophysics Source Code Library, record ascl:1109.001

\bibitem[{{Gonz{\'a}lez-Mart{\'\i}n} {et~al.}(2014){Gonz{\'a}lez-Mart{\'\i}n},
  {D{\'\i}az-Gonz{\'a}lez}, {Acosta-Pulido}, {Masegosa}, {Papadakis},
  {Rodr{\'\i}guez-Espinosa}, {M{\'a}rquez}, \&
  {Hern{\'a}ndez-Garc{\'\i}a}}]{Gonzalez2014}
{Gonz{\'a}lez-Mart{\'\i}n}, O., {D{\'\i}az-Gonz{\'a}lez}, D., {Acosta-Pulido},
  J.~A., {et~al.} 2014, \aap, 567, A92

\bibitem[{{Grassi} {et~al.}(2025){Grassi}, {Padovani}, {Galli}, {Vaytet},
  {Jensen}, {Redaelli}, {Spezzano}, {Bovino}, \& {Caselli}}]{Grassi2025}
{Grassi}, T., {Padovani}, M., {Galli}, D., {et~al.} 2025, arXiv e-prints,
  arXiv:2502.07874

\bibitem[{{Guiglion} {et~al.}(2024){Guiglion}, {Nepal}, {Chiappini},
  {Khoperskov}, {Traven}, {Queiroz}, {Steinmetz}, {Valentini}, {Fournier},
  {Vallenari}, {Youakim}, {Bergemann}, {M{\'e}sz{\'a}ros}, {Lucatello},
  {Sordo}, {Fabbro}, {Minchev}, {Tautvai{\v{s}}ien{\.{e}}}, {Mikolaitis}, \&
  {Montalb{\'a}n}}]{Guiglion2024}
{Guiglion}, G., {Nepal}, S., {Chiappini}, C., {et~al.} 2024, \aap, 682, A9

\bibitem[{{Heyl} {et~al.}(2023){Heyl}, {Butterworth}, \& {Viti}}]{Heyl2023}
{Heyl}, J., {Butterworth}, J., \& {Viti}, S. 2023, \mnras, 526, 404

\bibitem[{{Hinton} {et~al.}(2012){Hinton}, {Srivastava}, {Krizhevsky},
  {Sutskever}, \& {Salakhutdinov}}]{Hinton2012}
{Hinton}, G.~E., {Srivastava}, N., {Krizhevsky}, A., {Sutskever}, I., \&
  {Salakhutdinov}, R.~R. 2012, arXiv e-prints, arXiv:1207.0580

\bibitem[{{Humire} {et~al.}(2020){Humire}, {Thiel}, {Henkel}, {Belloche},
  {Loison}, {Pillai}, {Riquelme}, {Wakelam}, {Langer},
  {Hern{\'a}ndez-G{\'o}mez}, {Mauersberger}, \& {Menten}}]{Humire2020}
{Humire}, P.~K., {Thiel}, V., {Henkel}, C., {et~al.} 2020, \aap, 642, A222

\bibitem[{{Jaber Al-Edhari} {et~al.}(2017){Jaber Al-Edhari}, {Ceccarelli},
  {Kahane}, {Viti}, {Balucani}, {Caux}, {Faure}, {Lefloch}, {Lique}, {Mendoza},
  {Quenard}, \& {Wiesenfeld}}]{Al-Edhari2017}
{Jaber Al-Edhari}, A., {Ceccarelli}, C., {Kahane}, C., {et~al.} 2017, \aap,
  597, A40

\bibitem[{{Jimenez-Serra} {et~al.}(2025){Jimenez-Serra}, {Codella}, \&
  {Belloche}}]{Jimenez-Serra2025}
{Jimenez-Serra}, I., {Codella}, C., \& {Belloche}, A. 2025, arXiv e-prints,
  arXiv:2503.17104

\bibitem[{{J{\o}rgensen} {et~al.}(2020){J{\o}rgensen}, {Belloche}, \&
  {Garrod}}]{Jorgensen2020}
{J{\o}rgensen}, J.~K., {Belloche}, A., \& {Garrod}, R.~T. 2020, \araa, 58, 727

\bibitem[{{J{\o}rgensen} {et~al.}(2016){J{\o}rgensen}, {van der Wiel},
  {Coutens}, {Lykke}, {M{\"u}ller}, {van Dishoeck}, {Calcutt}, {Bjerkeli},
  {Bourke}, {Drozdovskaya}, {Favre}, {Fayolle}, {Garrod}, {Jacobsen},
  {{\"O}berg}, {Persson}, \& {Wampfler}}]{Jorgensen2016}
{J{\o}rgensen}, J.~K., {van der Wiel}, M.~H.~D., {Coutens}, A., {et~al.} 2016,
  \aap, 595, A117

\bibitem[{{Kami{\'n}ski} {et~al.}(2017){Kami{\'n}ski}, {Menten}, {Tylenda},
  {Karakas}, {Belloche}, \& {Patel}}]{Kaminski2017}
{Kami{\'n}ski}, T., {Menten}, K.~M., {Tylenda}, R., {et~al.} 2017, \aap, 607,
  A78

\bibitem[{{LeCun} {et~al.}(2015){LeCun}, {Bengio}, \& {Hinton}}]{LeCun2015}
{LeCun}, Y., {Bengio}, Y., \& {Hinton}, G. 2015, \nat, 521, 436

\bibitem[{{Lee} {et~al.}(2021){Lee}, {Patterson}, {Burkhardt}, {Vankayalapati},
  {McCarthy}, \& {McGuire}}]{Lee2021}
{Lee}, K. L.~K., {Patterson}, J., {Burkhardt}, A.~M., {et~al.} 2021, \apjl,
  917, L6

\bibitem[{{Lefloch} {et~al.}(2018){Lefloch}, {Bachiller}, {Ceccarelli},
  {Cernicharo}, {Codella}, {Fuente}, {Kahane}, {L{\'o}pez-Sepulcre}, {Tafalla},
  {Vastel}, {Caux}, {Gonz{\'a}lez-Garc{\'\i}a}, {Bianchi}, {G{\'o}mez-Ruiz},
  {Holdship}, {Mendoza}, {Ospina-Zamudio}, {Podio}, {Qu{\'e}nard}, {Roueff},
  {Sakai}, {Viti}, {Yamamoto}, {Yoshida}, {Favre}, {Monfredini},
  {Quiti{\'a}n-Lara}, {Marcelino}, {Boechat-Roberty}, \&
  {Cabrit}}]{Lefloch2018}
{Lefloch}, B., {Bachiller}, R., {Ceccarelli}, C., {et~al.} 2018, \mnras, 477,
  4792

\bibitem[{{Loomis} {et~al.}(2021){Loomis}, {Burkhardt}, {Shingledecker},
  {Charnley}, {Cordiner}, {Herbst}, {Kalenskii}, {Lee}, {Willis}, {Xue},
  {Remijan}, {McCarthy}, \& {McGuire}}]{Loomis2021}
{Loomis}, R.~A., {Burkhardt}, A.~M., {Shingledecker}, C.~N., {et~al.} 2021,
  Nature Astronomy, 5, 188

\bibitem[{{Loomis} {et~al.}(2018){Loomis}, {{\"O}berg}, {Andrews}, {Walsh},
  {Czekala}, {Huang}, \& {Rosenfeld}}]{Loomis2018}
{Loomis}, R.~A., {{\"O}berg}, K.~I., {Andrews}, S.~M., {et~al.} 2018, \aj, 155,
  182

\bibitem[{{Maret} {et~al.}(2011){Maret}, {Hily-Blant}, {Pety}, {Bardeau}, \&
  {Reynier}}]{Maret2011}
{Maret}, S., {Hily-Blant}, P., {Pety}, J., {Bardeau}, S., \& {Reynier}, E.
  2011, \aap, 526, A47

\bibitem[{{Mart{\'\i}n} {et~al.}(2021){Mart{\'\i}n}, {Mangum}, {Harada},
  {Costagliola}, {Sakamoto}, {Muller}, {Aladro}, {Tanaka}, {Yoshimura},
  {Nakanishi}, {Herrero-Illana}, {M{\"u}hle}, {Aalto}, {Behrens}, {Colzi},
  {Emig}, {Fuller}, {Garc{\'\i}a-Burillo}, {Greve}, {Henkel}, {Holdship},
  {Humire}, {Hunt}, {Izumi}, {Kohno}, {K{\"o}nig}, {Meier}, {Nakajima},
  {Nishimura}, {Padovani}, {Rivilla}, {Takano}, {van der Werf}, {Viti}, \&
  {Yan}}]{Martin2021}
{Mart{\'\i}n}, S., {Mangum}, J.~G., {Harada}, N., {et~al.} 2021, \aap, 656, A46

\bibitem[{{Mart{\'\i}n} {et~al.}(2019){Mart{\'\i}n}, {Mart{\'\i}n-Pintado},
  {Blanco-S{\'a}nchez}, {Rivilla}, {Rodr{\'\i}guez-Franco}, \&
  {Rico-Villas}}]{Martin2019}
{Mart{\'\i}n}, S., {Mart{\'\i}n-Pintado}, J., {Blanco-S{\'a}nchez}, C.,
  {et~al.} 2019, \aap, 631, A159

\bibitem[{{McGuire}(2022)}]{McGuire2022}
{McGuire}, B.~A. 2022, \apjs, 259, 30

\bibitem[{McGuire {et~al.}(2024)McGuire, Xue, Lee, El-Abd, \&
  Loomis}]{McGuire2024}
McGuire, B.~A., Xue, C., Lee, K. L.~K., El-Abd, S., \& Loomis, R.~A. 2024,
  molsim

\bibitem[{{Mendoza} {et~al.}(2025){Mendoza}, {Dall'Olio}, {Coelho},
  {Peregr{\'\i}n}, {L{\'o}pez-Dom{\'\i}nguez}, {van der Tak}, \&
  {Carvajal}}]{Mendoza2025}
{Mendoza}, E., {Dall'Olio}, P., {Coelho}, L.~S., {et~al.} 2025, \aap, 698, A286

\bibitem[{{Mercimek} {et~al.}(2022){Mercimek}, {Codella}, {Podio}, {Bianchi},
  {Chahine}, {Bouvier}, {L{\'o}pez-Sepulcre}, {Neri}, \&
  {Ceccarelli}}]{Mercimek2022}
{Mercimek}, S., {Codella}, C., {Podio}, L., {et~al.} 2022, \aap, 659, A67

\bibitem[{{Milam} {et~al.}(2005){Milam}, {Savage}, {Brewster}, {Ziurys}, \&
  {Wyckoff}}]{Milam2005}
{Milam}, S.~N., {Savage}, C., {Brewster}, M.~A., {Ziurys}, L.~M., \& {Wyckoff},
  S. 2005, \apj, 634, 1126

\bibitem[{{Mininni} {et~al.}(2020){Mininni}, {Beltr{\'a}n}, {Rivilla},
  {S{\'a}nchez-Monge}, {Fontani}, {M{\"o}ller}, {Cesaroni}, {Schilke}, {Viti},
  {Jim{\'e}nez-Serra}, {Colzi}, {Lorenzani}, \& {Testi}}]{Mininni2020}
{Mininni}, C., {Beltr{\'a}n}, M.~T., {Rivilla}, V.~M., {et~al.} 2020, \aap,
  644, A84

\bibitem[{{M{\"o}ller} {et~al.}(2013){M{\"o}ller}, {Bernst}, {Panoglou},
  {Muders}, {Ossenkopf}, {R{\"o}llig}, \& {Schilke}}]{Moller2013}
{M{\"o}ller}, T., {Bernst}, I., {Panoglou}, D., {et~al.} 2013, \aap, 549, A21

\bibitem[{{M{\"o}ller} {et~al.}(2025){M{\"o}ller}, {Schilke},
  {S{\'a}nchez-Monge}, \& {Schmiedeke}}]{Moller2025}
{M{\"o}ller}, T., {Schilke}, P., {S{\'a}nchez-Monge}, {\'A}., \& {Schmiedeke},
  A. 2025, \aap, 693, A160

\bibitem[{{Mookerjea} {et~al.}(2007){Mookerjea}, {Casper}, {Mundy}, \&
  {Looney}}]{Mookerjea2007}
{Mookerjea}, B., {Casper}, E., {Mundy}, L.~G., \& {Looney}, L.~W. 2007, \apj,
  659, 447

\bibitem[{{M{\"u}ller} {et~al.}(2005){M{\"u}ller}, {Schl{\"o}der}, {Stutzki},
  \& {Winnewisser}}]{Muller2005}
{M{\"u}ller}, H. S.~P., {Schl{\"o}der}, F., {Stutzki}, J., \& {Winnewisser}, G.
  2005, Journal of Molecular Structure, 742, 215

\bibitem[{{Möller} {et~al.}(2017){Möller}, {Endres}, \&
  {Schilke}}]{Moller2017}
{Möller}, T., {Endres}, C., \& {Schilke}, P. 2017, A\&A, 598, A7

\bibitem[{{Nazari} {et~al.}(2024){Nazari}, {Cheung}, {Asensio}, {Murillo}, {van
  Dishoeck}, {J{\o}rgensen}, {Bourke}, {Chuang}, {Drozdovskaya}, {Fedoseev},
  {Garrod}, {Ioppolo}, {Linnartz}, {McGuire}, {M{\"u}ller}, {Qasim}, \&
  {Wampfler}}]{Nazari2024}
{Nazari}, P., {Cheung}, J.~S.~Y., {Asensio}, J.~F., {et~al.} 2024, \aap, 686,
  A59

\bibitem[{{Nazari} {et~al.}(2022){Nazari}, {Meijerhof}, {van Gelder}, {Ahmadi},
  {van Dishoeck}, {Tabone}, {Langeroodi}, {Ligterink}, {Jaspers},
  {Beltr{\'a}n}, {Fuller}, {S{\'a}nchez-Monge}, \& {Schilke}}]{Nazari2022}
{Nazari}, P., {Meijerhof}, J.~D., {van Gelder}, M.~L., {et~al.} 2022, \aap,
  668, A109

\bibitem[{{Palau} {et~al.}(2017){Palau}, {Walsh}, {S{\'a}nchez-Monge},
  {Girart}, {Cesaroni}, {Jim{\'e}nez-Serra}, {Fuente}, {Zapata}, \&
  {Neri}}]{Palau2017}
{Palau}, A., {Walsh}, C., {S{\'a}nchez-Monge}, {\'A}., {et~al.} 2017, \mnras,
  467, 2723

\bibitem[{{Pan} \& {Yang}(2010)}]{Pan2010}
{Pan}, S.~J. \& {Yang}, Q. 2010, IEEE Transactions on Knowledge and Data
  Engineering, 22, 1345

\bibitem[{{Penzias}(1981)}]{Penzias1981}
{Penzias}, A.~A. 1981, \apj, 249, 518

\bibitem[{{Pickett} {et~al.}(1998){Pickett}, {Poynter}, {Cohen}, {Delitsky},
  {Pearson}, \& {M{\"u}ller}}]{Pickett1998}
{Pickett}, H.~M., {Poynter}, R.~L., {Cohen}, E.~A., {et~al.} 1998, \jqsrt, 60,
  883

\bibitem[{Qian(1999)}]{Qian1999}
Qian, N. 1999, Neural Networks, 12, 145

\bibitem[{{Qiu} {et~al.}(2025){Qiu}, {Zhang}, {M{\"o}ller}, {Jiang}, {Song},
  {Chen}, \& {Quan}}]{Qiu2025}
{Qiu}, Y., {Zhang}, T., {M{\"o}ller}, T., {et~al.} 2025, \apjs, 277, 21

\bibitem[{{Rolffs} {et~al.}(2011){Rolffs}, {Schilke}, {Zhang}, \&
  {Zapata}}]{Rolffs2011}
{Rolffs}, R., {Schilke}, P., {Zhang}, Q., \& {Zapata}, L. 2011, \aap, 536, A33

\bibitem[{{Sch{\"o}ier} {et~al.}(2005){Sch{\"o}ier}, {van der Tak}, {van
  Dishoeck}, \& {Black}}]{Schoier2005}
{Sch{\"o}ier}, F.~L., {van der Tak}, F.~F.~S., {van Dishoeck}, E.~F., \&
  {Black}, J.~H. 2005, \aap, 432, 369

\bibitem[{{Scolati} {et~al.}(2023){Scolati}, {Remijan}, {Herbst}, {McGuire}, \&
  {Lee}}]{Scolati2023}
{Scolati}, H.~N., {Remijan}, A.~J., {Herbst}, E., {McGuire}, B.~A., \& {Lee},
  K. L.~K. 2023, \apj, 959, 108

\bibitem[{{Sewi{\l}o} {et~al.}(2018){Sewi{\l}o}, {Indebetouw}, {Charnley},
  {Zahorecz}, {Oliveira}, {van Loon}, {Ward}, {Chen}, {Wiseman}, {Fukui},
  {Kawamura}, {Meixner}, {Onishi}, \& {Schilke}}]{Sewilo2018}
{Sewi{\l}o}, M., {Indebetouw}, R., {Charnley}, S.~B., {et~al.} 2018, \apjl,
  853, L19

\bibitem[{{Shimonishi} {et~al.}(2023){Shimonishi}, {Tanaka}, {Zhang}, \&
  {Furuya}}]{Shimonishi2023}
{Shimonishi}, T., {Tanaka}, K. E.~I., {Zhang}, Y., \& {Furuya}, K. 2023, \apjl,
  946, L41

\bibitem[{{Thiel} {et~al.}(2019){Thiel}, {Belloche}, {Menten}, {Giannetti},
  {Wiesemeyer}, {Winkel}, {Gratier}, {M{\"u}ller}, {Colombo}, \&
  {Garrod}}]{Thiel2019}
{Thiel}, V., {Belloche}, A., {Menten}, K.~M., {et~al.} 2019, \aap, 623, A68

\bibitem[{{Toru Shay} {et~al.}(2025){Toru Shay}, {Scolati}, {Wenzel}, {Lee},
  {Marimuthu}, \& {McGuire}}]{Toru2025}
{Toru Shay}, H., {Scolati}, H.~N., {Wenzel}, G., {et~al.} 2025, \apj, 985, 123

\bibitem[{{Vastel} {et~al.}(2015){Vastel}, {Bottinelli}, {Caux}, {Glorian}, \&
  {Boiziot}}]{Vastel2015}
{Vastel}, C., {Bottinelli}, S., {Caux}, E., {Glorian}, J.~M., \& {Boiziot}, M.
  2015, in SF2A-2015: Proceedings of the Annual meeting of the French Society
  of Astronomy and Astrophysics, ed. F.~{Martins}, S.~{Boissier}, V.~{Buat},
  L.~{Cambr{\'e}sy}, \& P.~{Petit}, 313--316

\bibitem[{{Vastel} {et~al.}(2024){Vastel}, {Sakai}, {Ceccarelli},
  {Jim{\'e}nez-Serra}, {Alves}, {Balucani}, {Bianchi}, {Bouvier}, {Caselli},
  {Chandler}, {Charnley}, {Codella}, {De Simone}, {Dulieu}, {Evans}, {Fontani},
  {Lefloch}, {Loinard}, {Menard}, {Podio}, {Sabatini}, {Sakai}, \&
  {Yamamoto}}]{Vastel2024}
{Vastel}, C., {Sakai}, T., {Ceccarelli}, C., {et~al.} 2024, \aap, 684, A189

\bibitem[{{Villadsen} {et~al.}(2022){Villadsen}, {Ligterink}, \&
  {Andersen}}]{Villadsen2022}
{Villadsen}, T., {Ligterink}, N.~F.~W., \& {Andersen}, M. 2022, \aap, 666, A45

\bibitem[{Wang {et~al.}(2025)Wang, Zhang, Bu, Lu, Duan, Quan, \&
  Qiu}]{Wang2025}
Wang, J., Zhang, Y., Bu, H., {et~al.} 2025, Intelligent Computing, 4, 0118

\bibitem[{{Widicus Weaver} \& {Friedel}(2012)}]{Widicus2012}
{Widicus Weaver}, S.~L. \& {Friedel}, D.~N. 2012, \apjs, 201, 16

\bibitem[{{Widicus Weaver} {et~al.}(2017){Widicus Weaver}, {Laas}, {Zou},
  {Kroll}, {Rad}, {Hays}, {Sanders}, {Lis}, {Cross}, {Wehres}, {McGuire}, \&
  {Sumner}}]{WidicusWeaver2017}
{Widicus Weaver}, S.~L., {Laas}, J.~C., {Zou}, L., {et~al.} 2017, \apjs, 232, 3

\bibitem[{{Wilson}(1999)}]{Wilson1999}
{Wilson}, T.~L. 1999, Reports on Progress in Physics, 62, 143

\bibitem[{{Wilson} \& {Rood}(1994)}]{Wilson1994}
{Wilson}, T.~L. \& {Rood}, R. 1994, \araa, 32, 191

\bibitem[{Wu \& He(2018)}]{Wu2018}
Wu, Y. \& He, K. 2018, Group Normalization

\bibitem[{{Yun} \& {Lee}(2023)}]{Yun2023}
{Yun}, H.-S. \& {Lee}, J.-E. 2023, \apj

\end{thebibliography}

\begin{acknowledgements}
The authors thank the referee for their valuable comments, which significantly improved the manuscript. This study received financial support from the French government in the framework of the University of Bordeaux's France 2030 program RRI Origins. N.~K. acknowledges financial support from the AAP Doctorat Intelligence Artificielle (U. Bordeaux, ANR AI). 
T.~Cs. received financial support from the French State in the framework of the IdEx Université de Bordeaux Investments for the future Program. This work was supported by the Programme National PCMI and PNPS of CNRS/INSU with INC/INP co-funded by CEA and CNES. Computer time for this study was provided by the computing facilities of the MCIA (Mésocentre de Calcul Intensif Aquitain).

\end{acknowledgements}

\begin{appendix}
    \onecolumn 

    \section{Complementary material} \label{appendix:complement_mol}
    
    Table \ref{tab:complement_isotop_vibration} lists the used  isotopologs and vibrationally excited states for the LTE models discussed in Sect.\,\ref{sub:models}, respectively. The used isotopic ratios are from \cite{Milam2005} ($^{12}$C/$^{13}$C = 68), \cite{Wilson1994} ($^{14}$N/$^{15}$N = 450), \cite{Penzias1981} ($^{16}$O/$^{17}$O = 2460), \cite{Wilson1999} ($^{16}$O/$^{18}$O = 560) and form \cite{Chin1996} ($^{32}$S/$^{33}$S = 153, $^{32}$S/$^{34}$S = 24).
    
    Table \ref{tab:small_molecules} gives detailed information on the species used to create the mask computed in Sect.\,\ref{sub:composite_spectra}. The used isotopic ratios are the same as previously mentioned. Table \ref{tab:hidden_transitions_mask} summarizes the proportion of blended transitions with this mask.
    
    Table \ref{tab:parameters_classical_hot_core} resumes the parameter range used for LTE modeling and the parameters used to produce a synthetic spectrum representative of a classical hot core. For the latter, the line width is set to $5$ \kms\ for all species and the emission is resolved (\ang{;;3} source for a \ang{;;3} beam). These parameters are compared to the $N_{\rm col}$ and $T_{\rm ex}$ measured for the hot cores Sgr~B2(N) \citep{Belloche2013}, G34.26+0.15 \citep{Mookerjea2007}, and the hot corino IRAS~16293B \citep{Nazari2024}.

    \begin{longtable}{c|c|c|c|c|c|c}
    \caption{\label{tab:complement_isotop_vibration} Summary of the isotopologs and the vibrational states used for LTE modeling}\\
    \hline\hline Main species & Isotopolog/State & Database & Tag & Version & Update  & Ratio to main species\\
    \hline
    CH$_3$CCH & $^{13}$CH$_3$CCH & CDMS & 41517 & V1 & 2020/04 & 68 \\
            & CH$_3^{13}$CCH & CDMS & 41516 & V1 & 2020/04 & 68 \\
            & CH$_3$C$^{13}$CH & CDMS & 41515 & V1 & 2020/04 & 68 \\
    CH$_3$CN & CH$_3^{13}$CN & CDMS & 42509 & V1 & 2017/08  & 68 \\
            & CH$_3$C$^{15}$N & CDMS & 42510 & V1 & 2017/08 & 450\\
    CH$_3$OH & $^{13}$CH$_3$OH & CDMS & 33502 & V1 & 2017/08 & 68\\
            & CH$_3^{18}$OH & CDMS & 34504 & V1 & 2017/08 & 560\\
    H$_2$CS & H$_2^{13}$CS & CDMS & 47505 & V2 & 2020/01 & 68 \\
        & H$_2$C$^{33}$S &  CDMS & 47506 & V2 & 2020/01 & 153\\
        & H$_2$C$^{34}$S &  CDMS & 48508 & V2 & 2020/01 & 24\\
    HC$_3$N & H$^{13}$CCCN & CDMS & 52509 & V1 & 2017/08 & 68\\
            & HC$^{13}$CCN & CDMS & 52510 & V1 & 2017/08  & 68\\
            & HCC$^{13}$CN & CDMS & 52511 & V1 & 2017/08  & 68\\
    HC(O)NH$_2$ & H$^{13}$C(O)NH$_2$ & CDMS & 46512 & V2 & 2017/08  & 68\\
    t-HCOOH & t-H$^{13}$COOH & CDMS & 47503 & V1 & 2015/06  & 68\\
    
    \hline 
    CH$_3$OH & $\varv 12 = 1$ & CDMS & 32504 & V3 & 2017/08 & 1 \\
            & $\varv 12 = 1 - 0$ & CDMS & 32505 & V3 & 2017/08 & 1\\
            & $\varv 12 = 2$ & JPL & 32003 & V3 & 2017/08 & 1\\
            & $\varv 12 = 2 - 0$ & JPL & 32003 & V3 & 2017/08 & 1\\
            & $\varv 12 = 2 - 1$ & JPL & 32003 & V3 & 2017/08 & 1\\
    $^{13}$CH$_3$OH & $\varv12 = 1$ & CDMS & 33502 & V1 & 2017/08 & 1\\
    CH$_3^{18}$OH & $\varv 12 = 1$ & CDMS & 34504 & V1 & 2017/08 & 1\\
            & $\varv 12 = 1 - 0$ & CDMS & 34504 & V1 & 2017/08 & 1\\
            & $\varv 12 = 2$ & CDMS & 34504 & V1 & 2017/08 & 1\\
            & $\varv 12 = 2 - 0$ & CDMS & 34504 & V1 & 2017/08 & 1\\
            & $\varv 12 = 2 - 1$ & CDMS & 34504 & V1 & 2017/08 & 1\\
     CH$_3$CN & $\varv 8 = 1$ & CDMS & 41509 & V1 & 2020/01 & 1\\
    \hline
    \end{longtable}

        \begin{longtable}{c|c|c|c|c|c|c|c}
    \caption{\label{tab:small_molecules} Summary of simple species and their number of transitions to mask.}\\
    \hline\hline
    Group & Molecule & Formula & Database & Tag & Version & Update & Nb. Transitions\\
    \hline 
    C-chains & Ethynyl radical & CCH & CDMS & 25501 & V3 & 2017/08 & 6 \\
    \hline O-bearing & Carbon monoxide & $^{13}$CO & CDMS & 29501 & V2 & 2017/08 & 1 \\
            &  & C$^{17}$O & JPL & 29006 & V2 & 2015/10 & 1\\ 
            &  & C$^{18}$O & CDMS & 30502 & V1 & 2017/08 & 1\\ 
             & Formyl cation & HCO$^+$ & CDMS & 29002 & V4 & 2019/12 & 4 \\
             & Formaldehyde & H$_2$CO & CDMS & 30501 & V3 & 2020/01 & 21 \\
             & Ketene & H$_2$CCO & CDMS & 42501 & V2 & 2017/08 & 40 \\
             & Silicon monoxide & SiO & CDMS &  44505 & V2 & 2017/08 & 1 \\
   \hline  N-bearing & Cyano radical & CN & CDMS & 26504 & V1 &  2017/08 & 9 \\
             & Hydrogen cyanide & HCN & CDMS & 27501 & V4 & 2017/08 & 1 \\
             & Hydrogen isocyanide & HNC & CDMS & 27502 & V2 & 2017/08 & 1 \\
             & Isocyanic acid & HNCO & CDMS & 43511 & V1 & 2017/08 & 50 \\
             & Diazenylium & N$_2$H$^+$ & CDMS & 29506 & V4 & 2017/08 & 1 \\
             & Hydroxylamine & NH$_2$OH & CDMS & 33503 & V1 & 2017/08 & 18 \\
   \hline S-bearing & Carbon monosulfide & CS & CDMS & 44501 & V2 & 2017/08 & 1 \\
             & Thioformyl cation & HCS$^+$ & CDMS & 45506 & V1 & 2017/08 & 1 \\
             & Hydrogen sulfide & H$_2$S & CDMS & 34502 & V1 & 2017/08 & 4 \\
             & Carbonyl sulfide & OCS & CDMS & 60503 & V2 & 2017/08 & 5 \\
             & Sulfur monoxide & SO & CDMS & 48501 & V1 & 2017/08 & 6 \\
             & Sulfur dioxide & SO$_2$ & CDMS & 64502 & V2 & 2017/08 & 110 \\
             \hline
    \end{longtable}

    \begin{longtable}{c|c|c}
    \caption{\label{tab:hidden_transitions_mask} Number of transitions of the modeled species and their percentage overlapping with the mask discussed in Sect.\,\ref{sub:composite_spectra}.} \\
    \hline \hline Molecule &  Masked transitions & Percentage of masked transitions \\ 
    \hline
$a$-(CH$_2$OH)$_2$ & 123 & 10.4 \% \\
C$_2$H$_3$CN & 28 & 13.0 \% \\
C$_2$H$_5$CN & 51 & 9.9 \% \\
C$_2$H$_5$OH & 203 & 9.1 \% \\
C$_3$H$_7$CN & 207 & 8.9 \% \\
CH$_3$CCH & 0 & 0.0 \% \\
CH$_3$CHO & 26 & 7.8 \% \\
CH$_3$CN & 9 & 23.7 \% \\
CH$_3$COCH$_3$ & 148 & 8.5 \% \\
CH$_3$NH$_2$ & 18 & 8.1 \% \\
CH$_3$OCH$_3$ & 30 & 6.9 \% \\
CH$_3$OCHO & 74 & 9.5 \% \\
CH$_3$OH & 43 & 8.1 \% \\
CH$_2$NH & 2 & 8.3 \% \\
$g$-(CH$_2$OH)$_2$ & 154 & 10.6 \% \\
HC$_3$N & 0 & 0.0 \% \\
HC(O)NH$_2$ & 22 & 9.6 \% \\
t-HCOOH & 14 & 7.3 \% \\
H$_2$CS & 5 & 6.1 \% \\
NH$_2$CN & 13 & 11.4 \% \\
       \hline
    \end{longtable}

\begingroup\small \setlength{\tabcolsep}{2pt}    
\begin{longtable}{c||c|c||c|c||c|c||c|c||c|c}
    \caption{\label{tab:parameters_classical_hot_core} Physical parameters used for the LTE models compared to literature values for hot core and hot corino sources.} \\
    \hline \hline Spectrum & \multicolumn{2}{c ||}{Parameter range} & \multicolumn{2}{c ||}{Synthetic hot core} & \multicolumn{2}{c ||}{Sgr B2(N)} & \multicolumn{2}{c ||}{IRAS16293B} & \multicolumn{2}{c }{G34.26+0.15}  \\
    & \multicolumn{2}{c ||}{Present study} & \multicolumn{2}{c ||}{Present study} & \multicolumn{2}{c ||}{\cite{Belloche2013}} & \multicolumn{2}{c ||}{\cite{Nazari2024}}  & \multicolumn{2}{c }{\cite{Mookerjea2007}} \\
    \hline Telescope & \multicolumn{2}{c ||}{ \ang{;;3} beam } & \multicolumn{2}{c ||}{ \ang{;;3} beam } &  \multicolumn{2}{c ||}{IRAM 30m} & \multicolumn{2}{c ||}{ALMA} & \multicolumn{2}{c }{BIMA}  \\
    \hline \hline Molecule &  $N_\text{col}$ [cm$^{-2}$]&  $T_\text{ex}$ [K] &  $N_\text{col}$ [cm$^{-2}$]&  $T_\text{ex}$ [K] &  $N_\text{col}$ [cm$^{-2}$]&  $T_\text{ex}$ [K] &  $N_\text{col}$ [cm$^{-2}$]&  $T_\text{ex}$ [K] &  $N_\text{col}$ [cm$^{-2}$]&  $T_\text{ex}$ [K]\\
    \hline
$a$-(CH$_2$OH)$_2$ & $10^{12}-10^{17}$ & $30 - 300 $ & $5\times10^{16}$  &   150 & $2.3\times10^{15}$    &   50   & $1.4\times10^{17}$  & 160  & - & - \\
   C$_2$H$_3$CN    & $10^{12}-10^{18}$ & $30 - 300 $ & $5\times10^{16}$  &   150 & $7.4\times10^{17}$    &   170  & $<4.0\times10^{15}$ & -    & - & \\
   C$_2$H$_5$CN    & $10^{12}-10^{18}$ & $30 - 300 $ & $10^{17}$         &   150 & $1.8\times10^{18}$    &   170  & $1.4\times10^{16}$  & 100  & $2.7\times10^{15}$  &  160 \\
   C$_2$H$_5$OH    & $10^{12}-10^{19}$ & $30 - 300 $ & $10^{17}$         &   150 & $9.1\times10^{17}$    &   150  & $6.2\times10^{16}$  & 170  & - & -  \\
   C$_3$H$_7$CN    & $10^{12}-10^{17}$ & $30 - 300 $ & $5\times10^{15}$  &   150 & $1.55\times10^{16}$   &   150  & -                   & -    & -                   & -   \\
      CH$_3$CCH    & $10^{12}-10^{19}$ & $30 - 300 $ & $10^{18}$         &    70 & $3.8\times10^{16}$    &    75  & -                   & -    & -                   & -   \\
      CH$_3$CHO    & $10^{12}-10^{19}$ & $30 - 300 $ & $10^{17}$         &   150 & $1.4\times10^{17}$    &   100  & $3.5\times10^{16}$  & 70   & - & - \\
       CH$_3$CN    & $10^{12}-10^{18}$ & $30 - 300 $ & $10^{17}$         &   150 & $2.0\times10^{18}$    &   200  & -                   & -    & $1.3\times10^{16}$  & 160   \\
 CH$_3$COCH$_3$    & $10^{12}-10^{19}$ & $30 - 300 $ & $10^{16}$         &   150 & $1.5\times10^{17}$    &   100  & $1.6\times10^{16}$  & 130  & - & -  \\
   CH$_3$NH$_2$    & $10^{12}-10^{19}$ & $30 - 300 $ & $5\times10^{17}$  &   150 & $6.0\times10^{17}$    &   100  & -                   & -    & -                   & -   \\
  CH$_3$OCH$_3$    & $10^{12}-10^{19}$ & $30 - 300 $ & $10^{17}$         &   150 & $2.0\times10^{18}$    &   130  & $8.5\times10^{16}$  & 100  & $5.7\times10^{16}$  &  160 \\
     CH$_3$OCHO    & $10^{12}-10^{19}$ & $30 - 300 $ & $10^{17}$         &   150 & $4.4\times10^{17}$    &   80   & $1.0\times10^{17}$  & 140  & - & - \\
       CH$_3$OH    & $10^{14}-10^{20}$ & $30 - 300 $ & $5\times10^{18}$  &   150 & $1.8\times10^{19}$    &   200  & $8.2\times10^{18}$  & -    & $3.4\times10^{17}$  & 160 \\
       CH$_2$NH    & $10^{12}-10^{19}$ & $30 - 300 $ & $10^{17}$         &   150 & $8.0\times10^{17}$    &   200  & -                   & -    & -                   & -   \\
$g$-(CH$_2$OH)$_2$ & $10^{12}-10^{17}$ & $30 - 300 $ & $10^{16}$         &   150 & -                     &   -    & $5.0\times10^{16}$  & 160  & -                   & -   \\
        HC$_3$N    & $10^{12}-10^{19}$ & $30 - 300 $ & $10^{15}$         &   150 & $1.2\times10^{16}$    &   60   & $6.2\times10^{13}$  & 100  & $5.1\times10^{13}$  &   160 \\
    HC(O)NH$_2$    & $10^{12}-10^{19}$ & $30 - 300 $ & $5\times10^{16}$  &   150 & $1.4\times10^{18}$    &   180  & $5.6\times10^{16}$  & 100  & $2.5\times10^{15}$  & 160 \\
        t-HCOOH    & $10^{12}-10^{19}$ & $30 - 300 $ & $5\times10^{16}$  &   70 & $1.55\times10^{16}$    &   70   & $7.7\times10^{16}$  & 300  & - & - \\
        H$_2$CS    & $10^{12}-10^{19}$ & $30 - 300 $ & $10^{17}$         &   150 & $2.5\times10^{17}$    &   150  & -                   & -    & -                   & -   \\
       NH$_2$CN    & $10^{12}-10^{19}$ & $30 - 300 $ & $10^{16}$         &   150 & $5.1\times10^{16}$    &   150  & -                   & -    & -                   & -   \\
        \hline
    
\end{longtable}
\endgroup
    
\section{Training dataset statistics}
\begin{figure*}[h]
    \centering
    \includegraphics[width=1\linewidth]{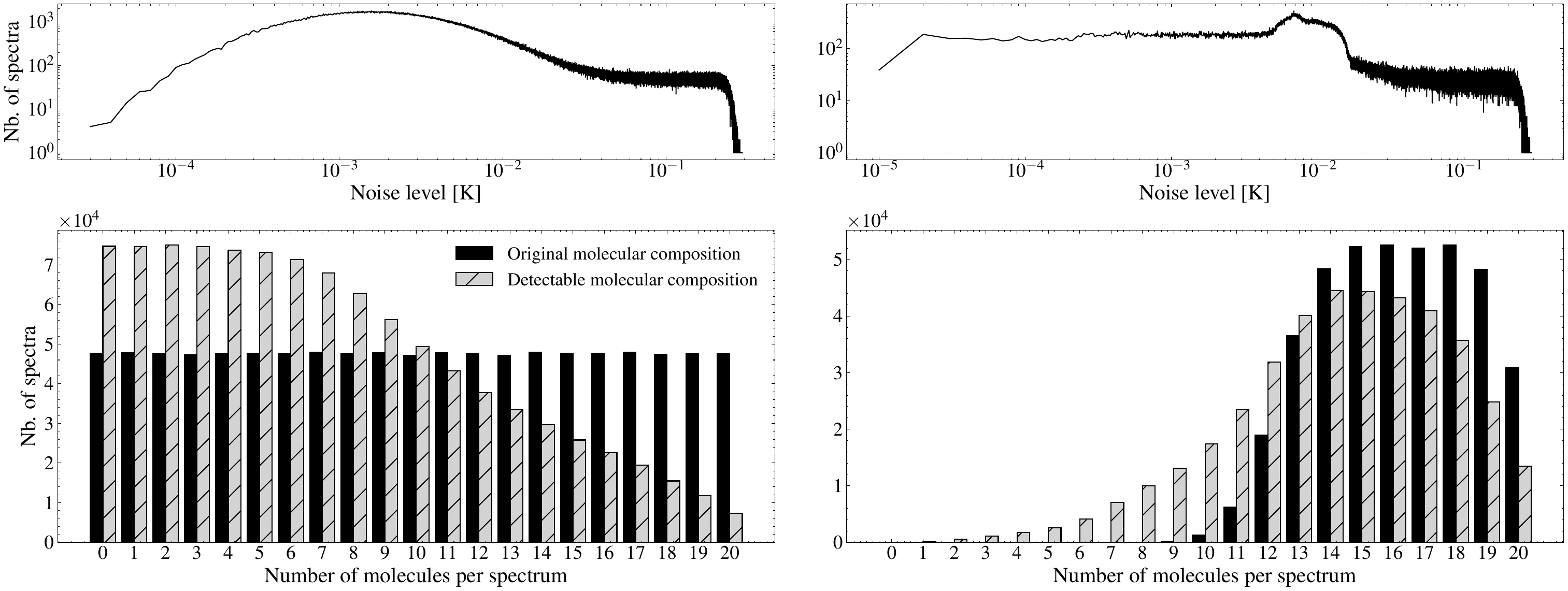}
    \caption{Histogram of the number of species in the composite synthetic spectra. The black bars correspond to the initial molecular composition, while gray dashed bars correspond to the number of detectable species after taking into account a varying noise distribution. \textbf{Left:} The "unconstrained" sub-set of synthetic spectra with a randomly selected molecular composition. \textbf{Right:} The "recipe" sub-set of data with a constrained molecular composition. The noise distribution is given on the top panel for each sub-set.}
    \label{fig:distrib_TS}
\end{figure*}

We show here the statistics of the "unconstrained" and "recipe" training sets. Fig. \ref{fig:distrib_TS} presents the distribution of the molecular content depending on the presence of the molecular spectra and the detectability of the molecular transitions in these spectra with respect to the noise level.

\section{Principle of the occlusion analysis} \label{appendix:occlusion_analysis}
\begin{figure}[h]
        \centering
        \includegraphics[width=0.7\linewidth]{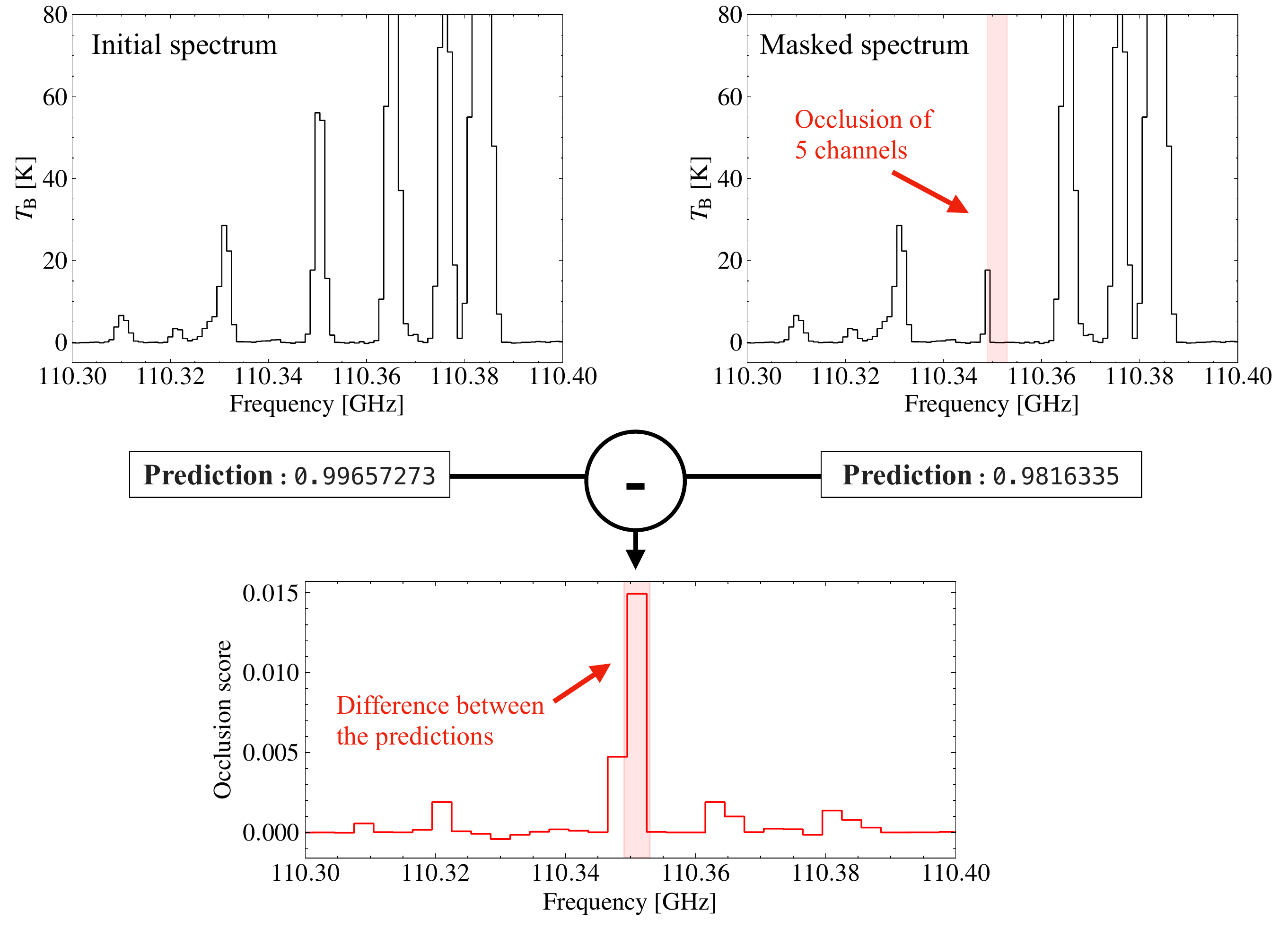}
        \caption{Principle of the occlusion analysis for the example of a \ce{CH3CN} LTE model.}
        \label{fig:principle_occlusion_analysis}
    \end{figure}

We use here the occlusion analysis as a model interpretation technique with the aim to assess the importance of specific channels in the CNN-model’s prediction. As discussed in Sect.\,\ref{sub:features} in detail, we perform a systematic masking (that we refer to as occlusion) small portions of the input spectrum using a sliding window. We then quantify the changes in the CNN-model’s output, where significant changes imply that the masked channels contain crucial information for its decision making. Figure \ref{fig:principle_occlusion_analysis} illustrates the principle of our occlusion analysis. The CNN-model generates a prediction for the original spectrum and for the modified version in which a small window of 5 channels is masked. The difference between these two predictions is computed and referred to as the occlusion score. This process is repeated across the entire spectrum.

Computing the occlusion score for each windowed region, we can map the relative contribution of the spectra to the prediction. Overall, this method provides us some interpretability by highlighting which frequency ranges the model relies on most heavily. In particular, comparing the frequency ranges used by the CNN-model with known molecular transitions allows us to assess whether its decision-making aligns with our expectation that it relies on physically meaningful spectral features.

\section{Data preparation and application of our CNN-model} \label{appendix:application}

\subsection{Preparation of the spectra}\label{app:prep_spectra}

The CNN-model can be easily applied to any new spectrum, regardless of the telescope specificity, resolution or setup. 
The CNN-model can also be adapted to new observational contexts, such as a different frequency coverage, different physical properties, or a different chemical richness through transfer learning (cf. Sect. \ref{sub:setups}). However, to use the CNN-model for a prediction, one must complete standard data reduction, such as continuum subtraction, \vlsr\ correction and resampling to the given frequency range with a 1~MHz spectral resolution. The latter can be done using a software such as PySpecKit \citep{Ginsburg2011}, CASSIS \citep{Vastel2015}, or CASA \citep{CASATeam2022}. Noticeable gaps in the spectra or artifacts should be put to 0.

\subsection{Importing the CNN-model and spectra to CIANNA}

The CNN-model is shared according to the Apache-2.0 license, with a mandatory reference to this article, and can be found at : \url{https://github.com/NinaK7/CNN-model-for-molecular-detection}. This CNN-model can be used or can be modified to be applied on new data. The test dataset (cf. Sect. \ref{sub:training_set}) and the "typical hot core spectrum" (cf. Sect. \ref{sub:composite_spectra}) mentioned in this article can be found on the GitHub web page.

Our CNN-model can be deployed in Python using the CIANNA API \citep{Cornu2025}. First, the spectra must be loaded that can be done from any format (ASCII or FITS), after having pre-formatted it as discussed  above in App.\,\ref{app:prep_spectra}. The input data need to be normalized according to the procedure presented in the paper, and converted to Numpy arrays compatible with the CIANNA format by the user before loading and running the trained model.

Model inference can be done on a lightweight hardware, no GPU is required for the application on a single spectrum, but the obtained performance may differ from the one from this article where we use a NVIDIA A100 GPU. In our situation, and at a computing precision of FP32C\_FP32A, the inference for $2\times10^4$ spectra made on the averaged CNN-model is done in less than 1.7 seconds, i.e., $\sim 1.2\times10^4$ spectra per second, and takes a RAM of 874 MB. The initialization of the framework in itself takes 0.30 s and the weight loading takes 0.28 s.

\end{appendix}

\end{document}